\documentclass[%
 aip,
 jmp,%
 amsmath,amssymb,
 reprint,%
]{revtex4-1}

\usepackage{amsmath}
\usepackage{graphicx}
\usepackage{xspace}
\usepackage{latexsym}
\usepackage{subfig}
\usepackage{braket}
\usepackage{cleveref}
\usepackage{threeparttable}
\usepackage{enumerate}
\usepackage{setspace}
\usepackage{mhchem}
\usepackage[dvipsnames]{xcolor}

\usepackage{array}
\newcolumntype{L}[1]{>{\raggedright\let\newline\\\arraybackslash\hspace{0pt}}m{#1}}
\newcolumntype{C}[1]{>{\centering\let\newline\\\arraybackslash\hspace{0pt}}m{#1}}
\newcolumntype{R}[1]{>{\raggedleft\let\newline\\\arraybackslash\hspace{0pt}}m{#1}}

\crefname{figure}{Figure}{Figures}
\crefname{table}{Table}{Tables}
\crefname{equation}{Eq.}{Eqs.}

\newcommand*{\degree}{\ensuremath{^\circ}\xspace}
\newcommand*{\meh}{\ensuremath{mE_h}\xspace}
\newcommand*{\eh}{\ensuremath{E_h}\xspace}

\newcommand{\h}[2]{h_{{#1}}^{{#2}}}
\renewcommand{\th}[2]{\tilde{h}_{{#1}}^{{#2}}}
\renewcommand{\d}[2]{\delta_{{#1}}^{{#2}}}
\newcommand{\gnobar}[2]{g_{{#1}}^{{#2}}}
\renewcommand{\v}[2]{{v}_{{#1}}^{{#2}}}
\renewcommand{\c}[1]{a^\dagger_{#1}}
\renewcommand{\a}[1]{a^{\ }_{#1}}
\newcommand{\f}[2]{f_{{#1}}^{{#2}}}
\newcommand{\e}[1]{\ensuremath{\varepsilon_{#1}}}
\newcommand{\pdm}[2]{\gamma_{{#1}}^{{#2}}}
\renewcommand{\AA}{\r{A}\xspace}

\newcommand{\eps}[2]{\epsilon_{{#1}}^{{#2}}}
\newcommand{\G}[2]{G_{{#1}}^{{#2}}}

\newcommand{\pc}{pc-NEVPT2\xspace}
\renewcommand{\sc}{sc-NEVPT2\xspace}
\newcommand{\tdpt}{t-NEVPT2\xspace}
\newcommand{\mrci}{MRCI+Q\xspace}
\newcommand{\caspt}{CASPT2\xspace}
\newcommand{\mad}{\ensuremath{\Delta_{\mathrm{MAD}}}\xspace}
\newcommand{\npe}{\ensuremath{\Delta_{\mathrm{NPE}}}\xspace}
\newcommand{\xsigma}{\ensuremath{X\,^{1}\Sigma_{g}^{+}}\xspace}
\newcommand{\bdelta}{\ensuremath{B\,^{1}\Delta_{g}}\xspace}
\newcommand{\bsigma}{\ensuremath{B^\prime\,^{1}\Sigma_{g}^{+}}\xspace}

\begin{document}

\raggedbottom 

\author{Alexander~Yu.~Sokolov}
\email{alexsokolov@princeton.edu}
\affiliation{Department of Chemistry, Princeton University, Princeton, New Jersey 08544, USA}
\author{Garnet~Kin-Lic~Chan}
\email{gkchan@princeton.edu}
\affiliation{Department of Chemistry, Princeton University, Princeton, New Jersey 08544, USA}
\title{A time-dependent formulation of multi-reference perturbation theory}

\begin{abstract}
  We discuss the time-dependent formulation of
  perturbation theory in the context of the interacting zeroth-order Hamiltonians
  that appear in multi-reference situations.
  As an example, we present  a time-dependent formulation and implementation of
  second-order $n$-electron valence perturbation theory.
  The resulting  \tdpt method yields the fully uncontracted $n$-electron valence
  perturbation wavefunction and energy, but has a lower computational scaling than the usual contracted variants,
  and also avoids the construction of high-order density matrices and
  the diagonalization of metrics. We present results of \tdpt for the water, nitrogen, carbon, and chromium molecules,
  and outline directions for the future.
\end{abstract}

\titlepage

\maketitle

\section{Introduction}

A standard strategy in the quantum chemistry of strongly correlated systems is to first compute
an approximate zeroth-order wavefunction within an active space of frontier molecular orbitals.
\cite{Olsen:1988p2185,Malmqvist:1990p5477,White:1999p4127,Legeza2008,Booth:2009p054106,Kurashige:2009p234114,Marti:2011p6750,Chan:2011p465,Wouters:2014p272}
The description of the remaining correlation outside of the active space is then the purview of multi-reference dynamic correlation theories.
\cite{Buenker:1974p33,Mukherjee:1977p955,Lindgren:1978p33,Siegbahn:1980p1647,Wolinski:1987p225,Hirao:1992p374,Finley:1998p299,Jeziorski:1981p1668,Mahapatra:1998p157,Mahapatra:1999p6171,Pittner:2003p10876,Evangelista:2007p024102,Werner:1988p5803,Andersson:1992p1218,Werner:1996p645,Angeli:2001p10252,Angeli:2001p297,Yanai:2006p194106,Yanai:2007p104107,Kurashige:2011p094104,Saitow:2013p044118,Saitow:2015p5120,Datta:2011p214116,Evangelista:2011p114102,Kohn:2012p176,Nooijen:2014p081102,Sharma:2014p111101,Sharma:2015p102815,Sokolov:2015p124107}
In principle, it is possible to directly extend single-reference dynamic correlation theories to the multi-reference setting.
However, several complications arise. First, the dimension of the space in which the perturbed first-order wavefunction resides is much larger
in the multi-reference case: $\sim \mathcal{O}(N_{det} \times N_{ext}^2)$, where $N_{det}$ is the dimension of the active Hilbert space,
and $N_{ext}$ is the number of external orbitals. Thus explicitly representing a general first-order wavefunction is costly. 
Second, the starting zeroth-order wavefunction is no longer an eigenstate of a one-electron Hamiltonian, but rather an interacting Hamiltonian.
This means that functions of the zeroth-order Hamiltonian, such as the resolvent operator in perturbation
theory, are not known in explicit computational form.
Third, the standard Wick's theorem, which reduces expectation values of fermionic operator strings
to products of single-particle density matrices, does not apply.\cite{Kutzelnigg:1997p432} This leads to
the need to compute expensive high-order density matrices.
It is common to circumvent these complications by introducing additional approximations which are not used
in the single-reference setting. Some of these are
internally contracted wavefunctions,\cite{Werner:1988p5803,Angeli:2001p10252,Angeli:2001p297} 
alternative zeroth-order Hamiltonians,\cite{Andersson:1992p1218,Li:2015p2097} and approximations to high-order density matrices.\cite{Yanai:2006p194106,Zgid:2009p194107,Saitow:2013p044118}
While enormously useful, these approximations can lead to new problems of their own, such as intruder states
associated with the choice of zeroth-order Hamiltonian, or metric instabilities associated with internal contraction.\cite{Evangelisti:1987p4930,Kowalski:2000p757,Kowalski:2000p052506,Evangelista:2014p054109}

In this work, we show that using a time-dependent formulation
ameliorates the complexity of the multi-reference dynamic correlation framework, without
the need to introduce additional approximations. 
As an example, we derive and implement the time-dependent $n$-electron valence second-order perturbation theory (\tdpt).
The resulting theory, although equivalent to the fully {\it uncontracted} $n$-electron valence perturbation theory,\cite{Angeli:2001p10252,Angeli:2001p297} exhibits a lower computational scaling
than the common contracted approximations for large active spaces, and avoids the need to diagonalize metric matrices
or compute four-particle density matrices.
Since the problem is time-independent, introducing time-dependence may also be viewed as a numerical trick
equivalent to the well-known Laplace transformation of perturbation theory.\cite{Almlof:1991p319,Haser:1992p489}
However, because the resolvent operator is not known explicitly for an interacting Hamiltonian, while time-evolution
with the same Hamiltonian is a relatively simple operation, the advantages of the Laplace transform are
greater in the multi-reference as compared to single-reference setting. 

In Section \ref{sec:Theory} we review time-independent and time-dependent many-body perturbation theory, taking note
of the relevant considerations for the multi-reference setting. In Section \ref{sec:implementation} we describe the efficient computational implementation of the time-dependent second-order $n$-electron valence perturbation theory.
In Sections \ref{sec:compdetails} and \ref{sec:results} we describe
the computational details and investigate the performance of \tdpt for a number of multi-reference problems with significant dynamic correlation: (i) the bond dissociation in \ce{H2O} and \ce{N2}; (ii) the ground and excited states in \ce{C2}; and (iii) the potential energy curve of the chromium dimer.
Finally, in Section \ref{sec:conclusions} we present our conclusions and future
outlook for the formulation.

\section{Theory}

\label{sec:Theory}
\subsection{Overview of time-independent perturbation theory}

\begin{figure}[!t]
  \includegraphics[width=0.25\textwidth]{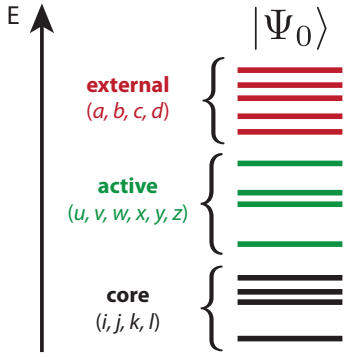}
   \captionsetup{justification=raggedright,singlelinecheck=false}
   \caption{Schematic orbital energy diagram showing the orbital index convention used for the active-space multi-reference wavefunction $\ket{\Psi_0}$.}
   \label{fig:mo_diagram}
\end{figure}

We begin with a brief overview of time-independent perturbation theory for multi-reference problems. We define the electronic Hamiltonian in second-quantized form as
\begin{align}
	\label{eq:hamiltonian}
	\hat{H} = \sum_{pq} \h{p}{q} \c{p}\a{q} + \frac{1}{4} \sum_{pqrs} \v{pq}{rs} \c{p}\c{q}\a{s}\a{r} \ ,
\end{align}
where $\h{p}{q}$ and $\v{pq}{rs}$ are the usual one- and antisymmetrized two-electron integrals:
\begin{align}
	\label{eq:integrals}
	\h{p}{q} 
	&= \braket{\psi_p(1)|\hat{h}|\psi_q(1)} \ , \quad 
	\v{pq}{rs} 
	= \gnobar{pq}{rs} - \gnobar{pq}{sr} \ , \\
	\gnobar{pq}{rs} 
	&= \braket{\psi_p(1)\psi_q(2)|\frac{1}{r_{12}}|\psi_r(1)\psi_s(2)} \ .
\end{align}
Indices $p,q,r,s$ run over the entire spin-orbital basis $\psi_p$. To
formulate multi-reference dynamic correlation theories, we partition the spin-orbitals into three sets: (i) {\it core} (doubly-occupied) with indices $i,j,k,l$; (ii) {\it active} with indices $u,v,w,x,y,z$; and (iii) {\it external} (unoccupied) with indices $a,b,c,d$ (\cref{fig:mo_diagram}). 

The Hamiltonian can be divided into two parts: 
\begin{align}
	\label{eq:H_parts}
	\hat{H} &= \hat{H}_{act} + \hat{H}_{nact} \ ,
\end{align}
where $\hat{H}_{nact}$ includes terms involving the core or external labels and $\hat{H}_{act}$ contains only the active-space contributions:
\begin{align}
	\label{eq:H_act}
	\hat{H}_{act} &= \sum_{xy}(\h{x}{y} + \sum_{i} \v{xi}{yi}) \c{x} \a{y}
	+ \frac{1}{4} \sum_{xywz} \v{xy}{zw} \c{x} \c{y} \a{w} \a{z} \ ,
\end{align}
where an additional term is included to describe interaction between the core and active electrons.

We now assume that we have determined a starting complete active-space (CAS) wavefunction $\ket{\Psi_0}$. 
To construct a multi-reference perturbation theory, we consider the wavefunction $\ket{\Psi_0}$ to be an eigenfunction of a zeroth-order Hamiltonian $\hat{H}^{(0)}$. While the choice of $\hat{H}^{(0)}$ is flexible, it is clear that it must be an interacting Hamiltonian. 
A convenient choice, which does not lead to intruder-state problems, and the one we will assume in later development, is the Dyall Hamiltonian\cite{Dyall:1995p4909} $\hat{H}^{(0)}=\hat{H}_D$, defined as:
\begin{align}
	\label{eq:H_Dyall}
	\hat{H}_D =& \sum_{ij} \f{i}{j} \c{i}\a{j} + \sum_{ab} \f{a}{b} \c{a}\a{b} + \hat{H}_{act} \ , \\
	\label{eq:f_gen}
	\f{p}{q} &= \h{p}{q} + \sum_{rs} \v{pr}{qs} \pdm{s}{r} \ ,
\end{align}
where $\pdm{q}{p}=\braket{\Psi_0|\c{p}\a{q}|\Psi_0}$ is the one-particle density matrix of $\ket{\Psi_0}$. Without loss of generality, we can work in the diagonal basis of the core and external generalized Fock operators, $\f{i}{j} \rightarrow \e{i} \d{i}{j}$, $\f{a}{b} \rightarrow \e{a} \d{a}{b}$, and henceforth we will assume that we do so.

We can now consider a direct expansion with respect to the perturbation $\lambda \hat{V}  = \lambda(\hat{H}-\hat{H}_D)$. For example, writing the ground-state energy of $\hat{H}_D + \lambda \hat{V}$ as $E(\lambda)$, we define the $n$th-order perturbation to the energy as:
\begin{align}
	\label{eq:E_n}
	E^{(n)} = \frac{1}{n!} \left.\frac{\partial^n E(\lambda)}{\partial \lambda^n}\right|_{\lambda=0} \ .
\end{align}
Starting from our choice of $\hat{H}_D$ as the reference Hamiltonian, the time-independent perturbation series corresponds to the fully uncontracted $n$-electron valence perturbation theory (NEVPT).\cite{Angeli:2001p10252,Angeli:2001p297} Denoting $\hat{V}^{\prime}$ as the part of $\hat{V}$ that contributes explicitly to the first-order wavefunction $\ket{\Psi^{(1)}}$ ($\hat{V}^{\prime} = \hat{Q} \hat{V}$, $\hat{Q} = 1 - \ket{\Psi_0}\bra{\Psi_0}$), the second-order energy can be written as:
\begin{align}
	\label{eq:E_tipt2}
	E^{(2)} &= \braket{\Psi_0 | \hat{V}^{\prime \dag}  | \Psi^{(1)}} \notag\\
	&= \braket{\Psi_0 | \hat{V}^{\prime\dag} \frac{1}{E_D - \hat{H}_D} \hat{V}^{\prime}|\Psi_0} \ .
\end{align}

We can now observe the  complications in
the uncontracted multi-reference perturbation theory described
above, which
arise in evaluating the expression \eqref{eq:E_tipt2}. The first-order wavefunction involves the resolvent $\frac{1}{E_D - \hat{H}_D}$, however, unlike in the single-reference setting,
the inversion appears to formally require inverting in the many-particle space. In addition, $\ket{\Psi^{(1)}}$ lives in a space of determinants
with $N-1$ or $N-2$ electrons in the core and active orbitals and $1$ or $2$ electrons in the external orbitals, respectively. Except
for very small basis sets, the
resulting number of determinants is
even larger than in the $N$-electron complete active space
used to determine $\ket{\Psi_0}$.
Based on these concerns, the above uncontracted multi-reference perturbation theory formulation is almost never used. Instead,
most implementations use an 
{\it internal contraction} approximation, where
$\ket{\Psi^{(1)}}$ is expanded in a set of contracted many-particle basis functions
as
\begin{align}
  \ket{\Psi^{(1)}} = \sum_I c_I^{(1)} \hat{O}_I \ket{\Psi_0} \ , \label{eq:contraction}
\end{align}
where $\hat{O}_I$ are operators which generate excitations. Different choices of $\hat{O}_I$ may be made, giving rise to, for example, the
strongly-contracted (sc-) and partially-contracted (pc-) variants of $n$-electron valence
perturbation theory.\cite{Angeli:2001p10252,Angeli:2001p297} Although the numerical error due to internal contraction is generally small,\cite{Werner:1988p5803} 
a significant drawback is that the basis states $\hat{O}_I \ket{\Psi_0}$ are no longer orthogonal. This introduces
a many-particle metric into the theory, which can either be diagonalized at large cost (e.g.\@ $\mathcal{O}(N_{act}^9)$, where $N_{act}$ is the number of active orbitals) or
which leads to numerical instabilities. Further, expectation values such
as the energy in Eq.~\eqref{eq:E_tipt2} involve the excitation
operators $\hat{O}_I$. These then require the evaluation of long strings of fermion operators and the computation of high-order density matrices.

Our thesis here is that, when considering larger active spaces, it can be more natural and computationally efficient
to work with the original uncontracted formulation, which is a direct extension of the single-reference formalism, rather
than with the more commonly implemented and approximate, contracted formulations. The key is to organize the uncontracted algorithm in an appropriate way. Such an organization is conveniently provided by the time-dependent formulation of the perturbation theory.

\subsection{Time-dependent perturbation theory}
We now consider multi-reference perturbation expansion from a time-dependent perspective. The general time-dependent perturbation theory\cite{March:1967} writes the ground-state wavefunction as:
\begin{align}
	\label{eq:WF_tdpt}
	\ket{\Psi(\lambda)} =
	\frac{\mathcal{T}
         \exp \left[-i \int_{-\infty}^0 \mathrm{d}t \ (\hat{H}^{(0)} + \lambda \hat{V}(t))\right]\ket{\Psi_0}}{\langle \Psi_0 | \mathcal{T} \exp \left[-i \int_{-\infty}^0 \mathrm{d}t \ (\hat{H}^{(0)} + \lambda \hat{V}(t))\right]\ket{\Psi_0}} \ ,
\end{align}
where the perturbation is switched on adiabatically at $t=-\infty$, i.e.  $\hat{V}(t)=\hat{V}e^{-\alpha|t|}$, and the operator $\mathcal{T}$
ensures time-ordering.

Since the perturbation $\hat{V}$ does not depend on time (aside from the exponentially vanishing adiabatic factor), the
time-dependent perturbation expansion \eqref{eq:WF_tdpt} must be identical to the time-independent
expansion discussed in Sec.\@ II A. For example, the $n$th-order wavefunction
$\ket{\Psi^{(n)}} = \frac{1}{n!} \frac{\partial^n \ket{\Psi(\lambda)}}{ \partial \lambda^n}$ is
identical to the $n$th-order wavefunction evaluated from the Rayleigh-Schr\"odinger
perturbation theory. Similarly, the $n$th-order energy given by
\begin{widetext}
\begin{align}
	\label{eq:E_tdpt}
	E^{(n)} 
	= i \partial_t \left. \left(
        \frac{(-i)^n}{n!}\int_{-\infty}^t \! \mathrm{d}t_1 \ldots \int_{-\infty}^t \! \mathrm{d}t_n \ \braket{\Psi_0 | \mathcal{T} \, \hat{V}_H(t_1) \hat{V}_H(t_2) \ldots \hat{V}_H(t_n) |\Psi_0}\right)_L \, \right|_{t=0} \ ,
\end{align}
\end{widetext}
where $\hat{V}_H(t) = e^{i(\hat{H}^{(0)}  - E^{(0)})t} \, \hat{V} \, e^{-i(\hat{H}^{(0)} - E^{(0)})t}$ and $L$ denotes linked contributions, is identical
to \cref{eq:E_n}.
Restricting our attention to second-order perturbation theory, \cref{eq:E_tdpt} can be written more simply as:
\begin{align}
	\label{eq:E_tdpt2}
	E^{(2)} 
        &= -\int_{0}^\infty \! \mathrm{d}\tau \braket{ \Psi_0 | \hat{V}^{\prime\dag}_H(\tau) \hat{V}^\prime_H(0) |\Psi_0} \notag \\
        &= -\int_{0}^\infty \! \mathrm{d}\tau \braket{ \Psi_0 | \hat{V}^{\prime\dag} e^{-(\hat{H}_D - E_D)\tau} \hat{V}^\prime |\Psi_0} \ , 
\end{align}
where only the $\hat{V}^\prime$ contributions are included, and we have
made the standard deformation of the integral from the real to imaginary $t$ axis with the substitution $\tau = it$ (Wick rotation).
Eq.~\eqref{eq:E_tdpt2} is no other than the well-known Laplace
transform expression of the second-order energy, and
is transparently equivalent to the time-independent result in \cref{eq:E_tipt2}. However, the presence of the reference Hamiltonian in the exponent rather than
in a denominator is a major advantage in the multi-reference setting. In particular, since the zeroth-order Hamiltonian for a multi-reference problem
must take the form
\begin{align}
\hat{H}^{(0)} = \hat{H}_{act} + \hat{H}_{core/ext} \ ,\label{eq:h0struct}
\end{align}
where $\hat{H}_{core/ext}$ and $\hat{H}_{act}$ commute, then the exponent factorizes as
\begin{align}
  e^{-\hat{H}^{(0)}\tau} = e^{-\hat{H}_{act} \tau} e^{-\hat{H}_{core/ext} \tau} \ ,
\end{align}
and the time-evolution can be carried out completely separately in the
active and core/external spaces. 
(Note that this does not require using the Dyall Hamiltonian specifically). 
 Because of this decoupling we do not need to consider the large $\mathcal{O}(N_{det} \times N_{ext}^2)$ dimension of the
first-order interacting space, and this removes the  barrier to working with
the uncontracted formulation
of multi-reference perturbation theory. Further, time-evolution with $\hat{H}_{act}$ 
is an operation of similar complexity to determining $\ket{\Psi_0}$ itself, and thus does not lead to an increase
in the computational scaling of the method. While it is possible, in principle, to use the structure of $\hat{H}^{(0)}$ without
referring to time-evolution, the time-dependent language provides a natural organization for  efficient algorithms. 

\subsection{Second-order time-dependent perturbation theory with Dyall Hamiltonian}

Let us now analyze the time-dependent formulation specializing to the choice of $\hat{H}^{(0)} = \hat{H}_D$.
We first analyze the structure of contributions to $E^{(2)}$ in \cref{eq:E_tdpt2}. It is convenient to divide $\hat{V}^\prime$ into 8 different terms: 
\begin{align}
	\label{eq:v_prime}
	\hat{V}^\prime 
	&= \hat{V}^{[0]} 
	+ \hat{V}^{[+1]} 
	+ \hat{V}^{[-1]} 
	+ \hat{V}^{[+2]} 
	+ \hat{V}^{[-2]} \\ \notag
	&+ \hat{V}^{[+1']} 
	+ \hat{V}^{[-1']} 
	+ \hat{V}^{[0']} \ ,
\end{align}
where the superscripts $[0]$, $[+1]$, $[-1]$, $\ldots$ are the conventional NEVPT2 notation,\cite{Angeli:2001p10252,Zgid:2009p194107} and do not denote the perturbation order. Explicit equations for each term of \cref{eq:v_prime} are given in the Appendix. Operators in \cref{eq:v_prime} give rise to 8 contributions to the second-order energy $E^{(2)}$:
\begin{align}
	\label{eq:e_contributions}
	E^{(2)} 
	&= E^{[0]} 
	+ E^{[+1]} 
	+ E^{[-1]} 
	+ E^{[+2]} 
	+ E^{[-2]} \\ \notag
	&+ E^{[+1']} 
	+ E^{[-1']} 
	+ E^{[0']}  \ .
\end{align}
As an example, we consider the $E^{[-2]}$ energy contribution arising from double excitations from active to external space ($xy\rightarrow ab$):
\begin{widetext}
\begin{align}
	E^{[-2]} 
	&= -\int_{0}^{\infty} \braket{ \Psi_0 | \hat{V}^{[-2]\dag}_H(\tau) \hat{V}^{[-2]}_H(0) |\Psi_0} \mathrm{d}\tau \notag \\
	&= -\frac{1}{16} \int_{0}^{\infty} \sum_{\substack{wxyz\\abcd}}\v{zw}{cd} \v{ab}{xy} \braket{ \Psi_0 | \c{z}(\tau)\c{w}(\tau)\a{d}(\tau)\a{c}(\tau) \c{a}\c{b}\a{y}\a{x} |\Psi_0} \mathrm{d}\tau \notag \\
	&= -\frac{1}{8} \int_{0}^{\infty} \sum_{\substack{wxyz\\ab}}\v{zw}{ab} \v{ab}{xy}  e^{-(\e{a} + \e{b}) \tau} \braket{ \Psi_0 | \c{z}(\tau)\c{w}(\tau)\a{y}\a{x} |\Psi_0} \mathrm{d}\tau \notag \\
	&= -\frac{1}{8} \int_{0}^{\infty} \sum_{wxyz} \eps{zw}{xy}(\tau) \G{xy}{zw}(\tau) \, \mathrm{d}\tau \ ,
	\label{eq:e_-2_deriv}
\end{align}
\end{widetext}
where we defined an intermediate $\eps{zw}{xy}(\tau)=\sum_{ab}\v{zw}{ab} \v{ab}{xy}  e^{-(\e{a} + \e{b})\tau}$ and used the property of the time-dependence of the Dyall Hamiltonian in the external space: $\a{a}(\tau) = e^{-\e{a} \tau} \a{a}(0) = e^{-\e{a} \tau} \a{a}$. In \cref{eq:e_-2_deriv}, $\G{xy}{zw}(\tau)=\braket{ \Psi_0 | \c{z}(\tau)\c{w}(\tau)\a{y}\a{x} |\Psi_0}$ is a two-particle one-time Green's function of the Dyall Hamiltonian in the active space. In general, the time-ordered $m$-particle $n$-time Green's function is defined in imaginary time as:
\begin{align}
	\label{eq:gf_general}
	G^{pq \ldots}_{rs \ldots}(\tau_1, \tau_2, \ldots \tau_n) 
	= \braket{\Psi_0| \mathcal{T} \c{p}(\tau_1) \c{q}(\tau_2) \ldots \a{r} (\tau_n) |\Psi_0} \ .
\end{align}
The highest rank active-space Green's function that
appears in \cref{eq:E_tdpt2} arises from the $\hat{V}^\prime$ contributions that involve 3 active-space labels, namely $\hat{V}^{[-1']}$ and $\hat{V}^{[+1']}$. Contractions of these operators yield three-particle one-time Green's functions $\braket{\Psi_0| \c{u}(\tau) \c{v}(\tau) \a{w}(\tau) \c{z} \a{y} \a{x} |\Psi_0}$ and $\braket{\Psi_0| \c{u}(\tau) \a{v}(\tau) \a{w}(\tau) \c{z} \c{y} \a{x} |\Psi_0}$, respectively, as shown in the Appendix.
 
The one-, two-, and three-particle, one-time, active-space Green's functions are the central objects to compute in time-dependent second-order $n$-electron
valence perturbation theory (\tdpt). The reduction of the energy to these compact quantities is an important strength of the time-dependent view. Physically, this simplicity arises because electron correlation effects are implicitly described by the time-dependence of the Green's functions. 
We stress that although the time-dependent second-order energy is expressed entirely in terms of reduced Green's functions, reminiscent of the internal contraction approximation, there is no approximation being made -- \cref{eq:E_tdpt2} yields the fully uncontracted theory. Thus, there is no inversion of a metric tensor, which is a bottleneck for large active spaces and basis sets in internally-contracted theories, and the most complicated object to appear is the three-particle Green's function, in contrast to the four-particle density matrix in time-independent internally-contracted NEVPT2.
Additional costs arise from the need to compute a time-evolution and time-integration in \cref{eq:E_tdpt2}, however, as we will demonstrate in Sec.\@ III, these tasks are not computationally difficult and can be carried out very efficiently.

\section{Implementation}
\label{sec:implementation}

We now briefly discuss the implementation of \tdpt for complete active-space self-consistent field (CASSCF) reference wavefunctions.\cite{Shepard:1987p63,Roos:1987p399,Olsen:1988p2185} The \tdpt algorithm is summarized below and can be easily implemented using any available full configuration interaction (FCI) or CASSCF computer program. The individual steps are:
\begin{enumerate}
\item Compute initial active-space wavefunctions at $\tau = 0$ (e.g., $\ket{\Psi^x}=\c{x}\ket{\Psi_0}$, $\ket{\Psi_{xy}}=\a{x}\a{y}\ket{\Psi_0}$, etc). There are seven types of active-space states to be computed, namely $\ket{\Psi^{x}}$, $\ket{\Psi_{x}}$, $\ket{\Psi^{xy}}$, $\ket{\Psi_{xy}}$, $\ket{\Psi^{x}_{y}}$, $\ket{\Psi^{xy}_{z}}$, and $\ket{\Psi^{x}_{yz}}$ (see Appendix for details). 

\item Loop over time steps starting with $\tau=0$. Compute reduced Green's functions by evaluating the overlap matrix elements $G(\tau')=\braket{\Phi(\tau')|\Phi}$, where $\ket{\Phi(\tau')}$ and $\ket{\Phi}$ are the active-space states ($\ket{\Psi^{x}}$, $\ket{\Psi^{xy}}$, $\ldots$) at $\tau = \tau'$ and $\tau = 0$, respectively. For example, for $\ket{\Phi} = \ket{\Psi_{yx}}$, the corresponding two-particle Green's function is $G(\tau') = \braket{\Psi_{wz}(\tau')|\Psi_{yx}}=\braket{\Psi_0|\c{z}(\tau')\c{w}(\tau')\a{y}\a{x}|\Psi_0}$. 

\item Compute correlation energy contributions at $\tau=\tau'$ as the product of one- and two-electron integrals and the active-space Green's functions: $E(\tau') = \eps{}{}(\tau')G(\tau')$. For example, for $\ket{\Phi} = \ket{\Psi_{yx}}$, the integral prefactor $\eps{}{}(\tau')=\eps{zw}{xy}(\tau')=-\frac{1}{8}\sum_{ab}\v{zw}{ab} \v{ab}{xy} e^{-(\e{a} + \e{b})\tau'}$ and the corresponding energy contribution $E^{[-2]}(\tau')= \sum_{wxyz} \eps{zw}{xy}(\tau') \braket{\Psi_{wz}(\tau')|\Psi_{yx}}$.

\item If the magnitude of the correlation energy at $\tau=\tau'$ is less than the energy convergence threshold ($|E(\tau')| < \Delta E_{conv}$), proceed to step 5. If $|E(\tau')| > \Delta E_{conv}$, propagate the active-space states $\ket{\Phi}$ according to the time-dependent Schr\"odinger equation:
\begin{align}
	\label{eq:propagate}
	\ket{\Phi(\tau)} = e^{-(\hat{H}_D - E_D) \tau}  \ket{\Phi} \ , 
\end{align} 
where $\tau = \tau' + \delta \tau$ and $\delta \tau$ is a step in imaginary time. Return to step 2.

\item Compute correlation energy by time-integration: 
\begin{align}
	\label{eq:integrate}
		E^{(2)} = \int_{0}^{\infty} \! E(\tau) \, \mathrm{d}\tau  \ .
\end{align} 
\cref{eq:integrate} is evaluated by fitting the computed values $E(\tau)$ to an exponential function $\mathcal{E}(\tau) = \sum_i a_i e^{-b_i \tau}$, followed by the analytic integration of the obtained result. Depending on the desired accuracy, fitting typically requires a linear combination of 6 to 12 exponentials.
\end{enumerate}

The basic \tdpt algorithm outlined above has $\mathcal{O}(N_\tau \times N_{det} \times N_{act}^7)$ computational cost, where $N_{det}$ is the dimension of the active-space Hilbert space, $N_\tau$ is the number of time steps, and $N_{act}$ is the number of active orbitals. The efficiency of the algorithm
is greatly improved if we avoid the explicit computation and storage of the $\ket{\Psi^{xy}_{z}}$ and $\ket{\Psi^{x}_{yz}}$ states by contracting these wavefunctions with the corresponding two-electron integral tensors. For example, the three-particle contribution to $E^{[-1']}$ can be evaluated by computing a vector
\begin{align}
	\label{eq:v_a}
	\ket{v_a} = \sum_{xyz}\v{ax}{zy} \c{x} \a{y} \a{z} \ket{\Psi_0} \ ,
\end{align}
propagating $\ket{v_a(\tau)} = e^{-(\hat{H}_D - E_D) \tau} \ket{v_a}$, and evaluating the integral:
\begin{align}
	\label{eq:v_a_energy}
	E^{[-1']} \Leftarrow - \frac{1}{4} \int_{0}^{\infty}  \sum_{a} e^{-\e{a} \tau} \braket{v_a(\tau)|v_a} \mathrm{d}\tau \ .
\end{align}
Avoiding the explicit time-propagation of the $\ket{\Psi^{xy}_{z}}$ and $\ket{\Psi^{x}_{yz}}$ states lowers the computational scaling of the \tdpt algorithm to $\mathcal{O}(N_\tau \times N_{det} \times N_{ext} \times N_{act}^4 )$ + $\mathcal{O}(N_\tau \times N_{det} \times N_{act}^6)$, where $N_{ext}$ is the number of external orbitals. This is significantly less than the cost of computing the four-particle density matrix in internally-contracted time-independent theories such as sc- and pc-NEVPT2 ($\mathcal{O} (N_{det} \times N_{act}^8)$), when using large active spaces. 

\begin{figure*}[t]
   \subfloat[]{\label{fig:algorithm_errors_2}\includegraphics[width=0.45\textwidth]{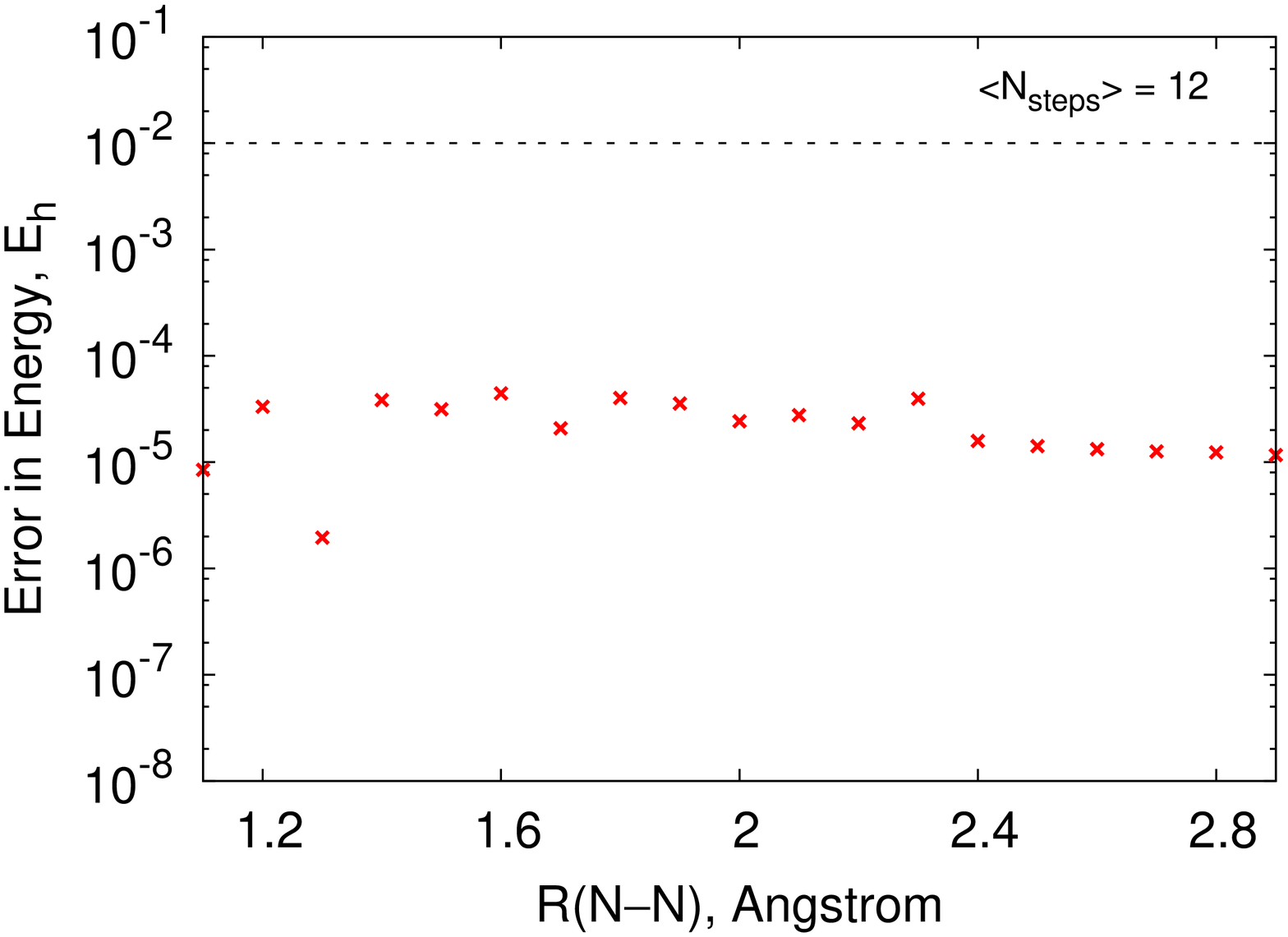}} \qquad
   \subfloat[]{\label{fig:algorithm_errors_3}\includegraphics[width=0.45\textwidth]{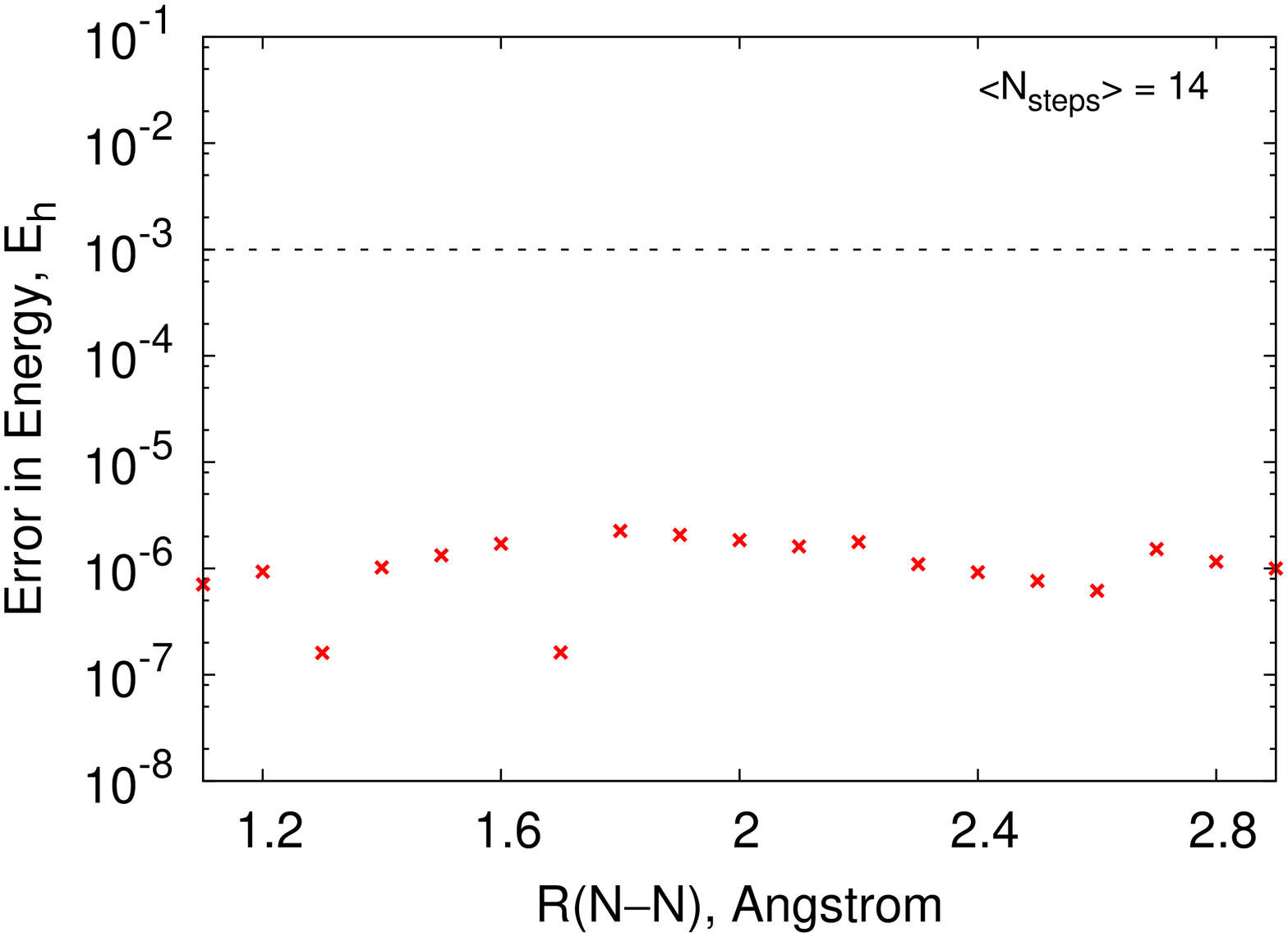}} \quad
   \subfloat[]{\label{fig:algorithm_errors_4}\includegraphics[width=0.45\textwidth]{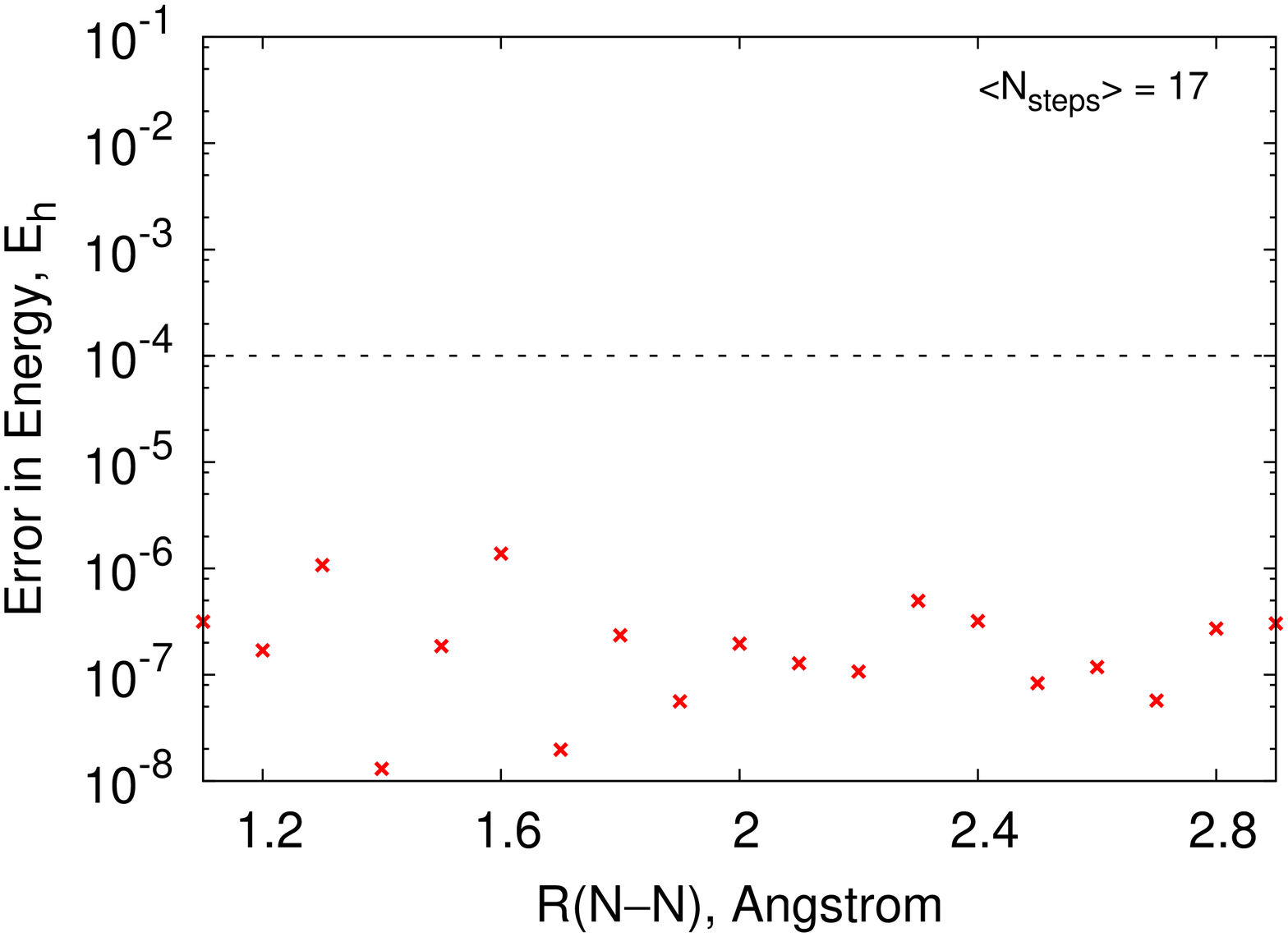}} \qquad
   \subfloat[]{\label{fig:algorithm_errors_5}\includegraphics[width=0.45\textwidth]{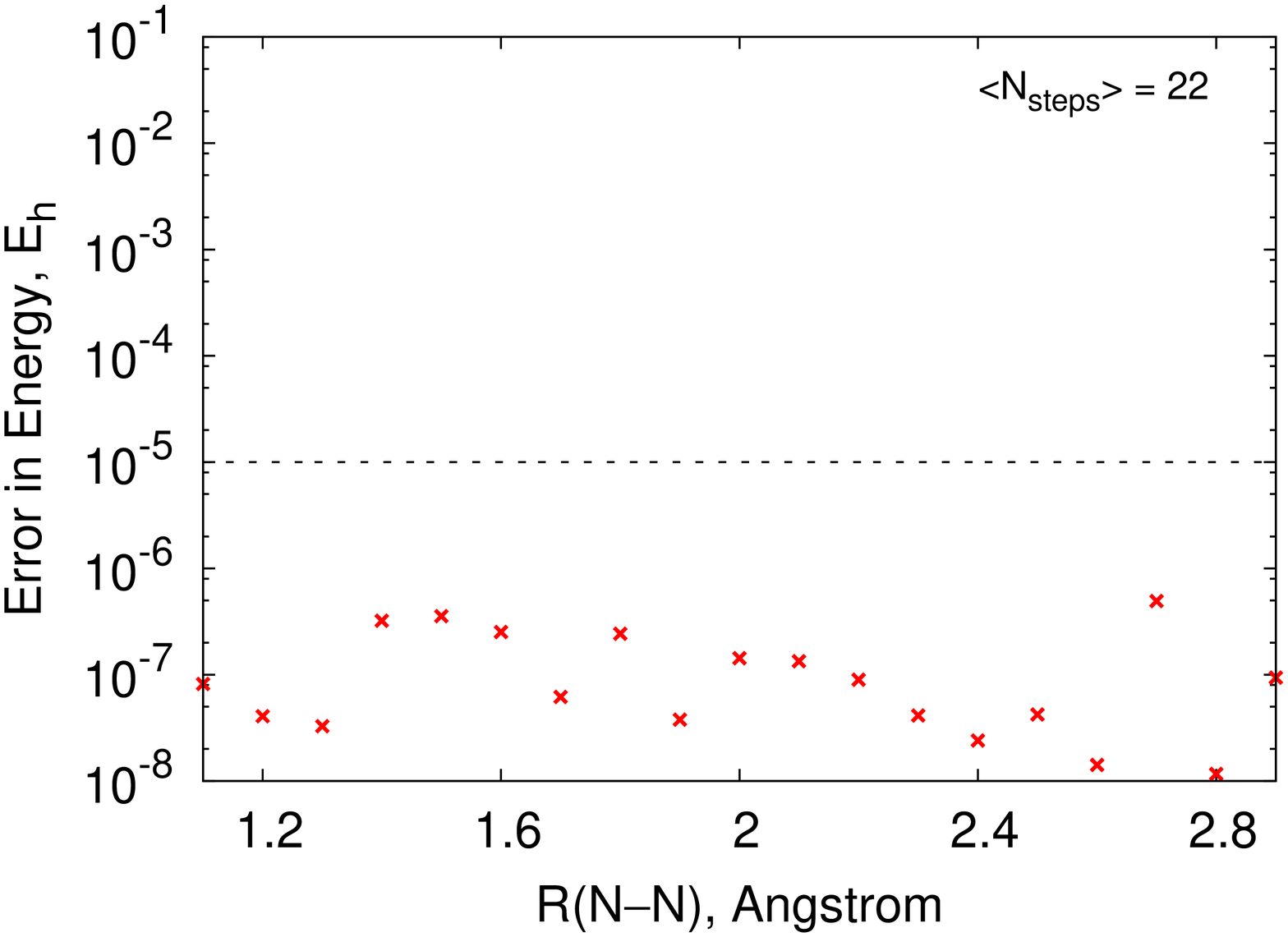}} 
   \captionsetup{justification=raggedright,singlelinecheck=false}
   \caption{Errors of the \tdpt correlation energy (in \eh), relative to the uncontracted NEVPT2 correlation energy, computed for different values of the $\Delta E_{conv}$ parameter (see Sec.\@ \ref{sec:implementation} for details). The error values are plotted as red crosses for a range of \ce{N2} bond distances. The $\Delta E_{conv}$ values are depicted as horizontal dashed lines. Computations employed the (6e, 6o) CASSCF reference wavefunction and the \mbox{6-311G} basis set. Also shown is the average number of time steps taken during the time-propagation ($\langle N_{steps}\rangle$).}
   \label{fig:algorithm_errors}
\end{figure*}

For further efficiency improvements, it is desirable to reduce the number of time steps $N_\tau$ while maintaining the accuracy of time-propagation and time-integration. In our implementation, the time-propagation in \cref{eq:propagate} is performed using the embedded Runge-Kutta (ERK) (4,5) algorithm,\cite{Press:2007} which automatically determines the time step $\delta \tau$ to
use in a fourth-order Runge-Kutta propagation based on the error estimate of the fifth-order propagator approximation, as well as the specified convergence threshold $\Delta E_{conv}$. We begin time-propagation by setting $\delta \tau$ to a small value ($\sim 10^{-3}$). At each iteration, the new time step is determined as $\delta \tau' = \mathrm{min}(2\times\delta \tau, \delta\tau_{emb})$, where $\delta\tau_{emb}$ is the time step estimated using the ERK method. The resulting algorithm typically requires 10-15 time-propagation steps to achieve 0.1 \meh accuracy in the correlation energy $E^{(2)}$. \cref{fig:algorithm_errors} demonstrates the errors in $E^{(2)}$ for the dissociation of the \ce{N2} molecule, relative to the exact uncontracted NEVPT2 correlation energy computed, for different values of $\Delta E_{conv}$. 

Finally, we stress that each step of the \tdpt algorithm can be efficiently parallelized. In particular, computation, memory storage and time-propagation of the active-space states $\ket{\Phi}$ can be performed in an ``embarrassingly'' parallel fashion, with zero communication between processors. In our parallel implementation, each processor is assigned to compute a subset of states $\{\ket{\Phi}\}_{i}$ ($i=1,\ldots,K$), where $K$ is the total number of processors (step 1 of the algorithm above). At every time step $\tau=\tau'$, the $i$th subset $\{\ket{\Phi(\tau')}\}_{i}$ is propagated on the $i$th processor (step 4). The correlation energy contributions from each subset $\{\ket{\Phi(\tau')}\}_{i}$ can be computed by evaluating the Green's function matrix elements $G(\tau')=\braket{\Phi(\tau')|\Phi}_{i}$, where the states $\ket{\Phi}$ at $\tau = 0$ are either recomputed or accessed from memory (steps 2 and 3). We do not exploit parallelism for the energy evaluation in our current implementation, since this is a relatively inexpensive step of the algorithm.

\section{Computational Details}
\label{sec:compdetails}
\tdpt was implemented as a standalone Python code interfaced with the \textsc{Pyscf} program.\cite{PYSCF} For details about the \tdpt implementation, see Sec.\@ \ref{sec:implementation}. Our \tdpt code reproduces the uncontracted NEVPT2 results using the MPS-PT2 algorithm reported by Sharma and Chan.\cite{Sharma:2014p111101} All multi-reference methods employed the complete active-space self-consistent field (CASSCF) wavefunctions\cite{Shepard:1987p63,Roos:1987p399,Olsen:1988p2185} as the reference, with active spaces of $n$ electrons in $m$ orbitals denoted as ($n$e, $m$o). Computations using the complete active-space second-order perturbation theory (CASPT2),\cite{Andersson:1992p1218} partially-contracted $n$-electron valence second-order perturbation theory (pc-NEVPT2),\cite{Angeli:2001p10252,Angeli:2001p297} as well as multi-reference configuration interaction theory with single and double excitations (MRCI)\cite{Werner:1988p5803} were performed using the \textsc{Molpro} package.\cite{MOLPRO} The MRCI energies were supplied with the Davidson correction; the resulting method is denoted as MRCI+Q.\cite{Langhoff:1974p61,Meissner:1988p204} In all computations, the cc-pV$X$Z ($X$ = D, T, Q, 5) basis sets\cite{Dunning:1989p1007} were used, unless noted otherwise. For the chromium dimer, the \tdpt correlation energies were converged to $\Delta E_{conv}$ = $10^{-4}$ \eh, while for all other systems the tighter $\Delta E_{conv}$ = $10^{-5}$ \eh parameter was used. The \ce{Cr2} total energies were extrapolated to the complete basis set limit (CBS) using the following equations:\cite{Feller:1993p7059,Helgaker:1997p9639}
\begin{align}
	\label{eq:e_cas_ext}
	E_{\mathrm{ref}} (X) =& E_{\mathrm{ref}}^{\mathrm{\mathrm{CBS}}} + Ae^{-B(X+1)}\ , \\
	E_{\mathrm{corr}} (X) =& E_{\mathrm{corr}}^{\mathrm{CBS}} + C(X+1)^{-3}\ ,
\end{align}
where $E_{\mathrm{ref}} (X)$ and $E_{\mathrm{corr}} (X)$ are the CASSCF reference and the corresponding correlation energies, respectively, computed using the cc-pV$X$Z basis sets. To obtain $E_{\mathrm{corr}}^{\mathrm{CBS}}$, the $E_{\mathrm{corr}}(X)$ ($X$ = Q, 5) correlation energies were used.

\section{Results}\label{sec:results}
\subsection{Covalent bond dissociation in \ce{H2O} and \ce{N2}}
\begin{figure}[t]
   \includegraphics[width=0.45\textwidth]{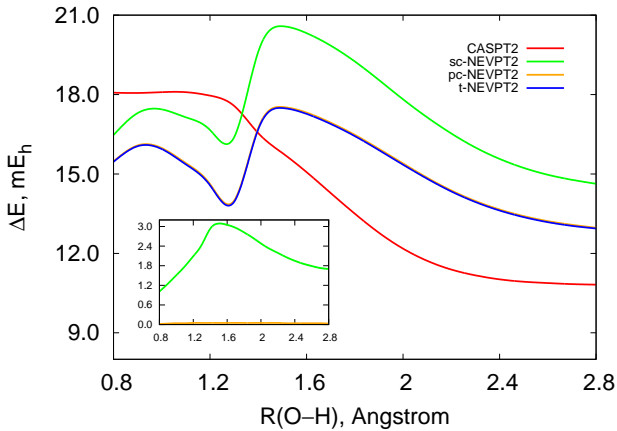}
   \captionsetup{justification=raggedright,singlelinecheck=false}
   \caption{Energy (in \meh) as a function of the O--H bond length for the symmetric bond dissociation of water, relative to multi-reference configuration interaction (\mrci). All methods employed the (6e, 9o) CASSCF reference wavefunction and the \mbox{cc-pVQZ} basis set. The H--O--H angle was fixed at 104.5\degree. The inset shows deviation of the strongly and partially-contracted NEVPT2 energies from the uncontracted \tdpt results.}
   \label{fig:H2O}
\end{figure}

\begin{figure*}[t]
   \subfloat[]{\label{fig:N2_full}\includegraphics[width=0.45\textwidth]{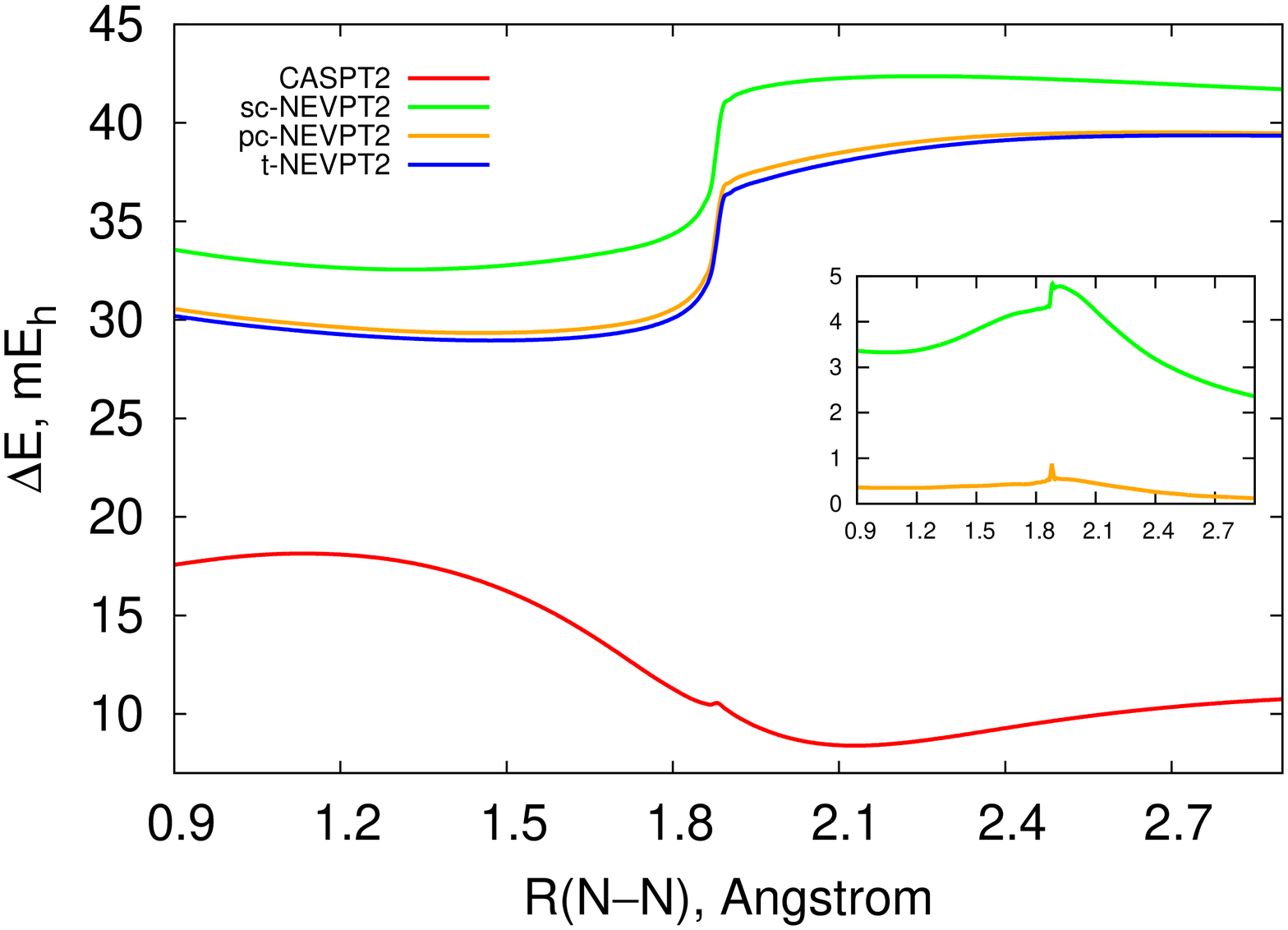}} \qquad
   \subfloat[]{\label{fig:N2_zoomed}\includegraphics[width=0.45\textwidth]{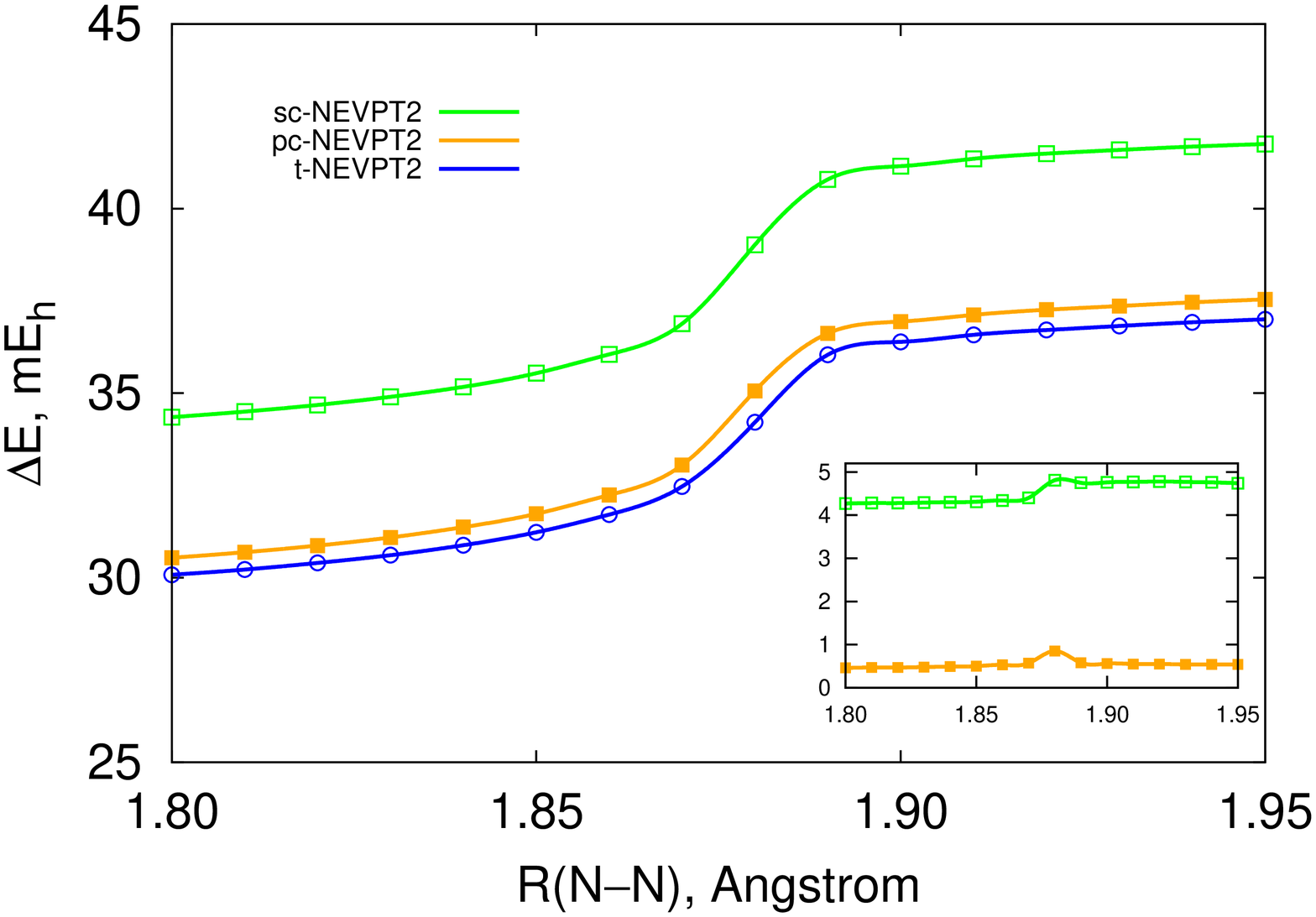}}
   \captionsetup{justification=raggedright,singlelinecheck=false}
   \caption{Energy (in \meh) as a function of the N--N bond length for the dissociation of \ce{N2}, relative to multi-reference configuration interaction (\mrci). All methods employed the (10e, 10o) CASSCF reference wavefunction and the \mbox{cc-pVQZ} basis set. Plot (a) shows results from the equilibrium region to dissociation, while (b) presents a detailed plot for the \mbox{1.80 -- 1.95 \AA} region. In both plots, the inset shows deviation of the strongly and partially-contracted NEVPT2 energies from the uncontracted \tdpt results. }
   \label{fig:N2}
\end{figure*}

We begin by investigating the accuracy of \tdpt for the description of the bond dissociation of \ce{N2} and the symmetric bond stretching of \ce{H2O}. \cref{tab:H2O,tab:N2} present the total energies of \ce{H2O} and \ce{N2} computed using \tdpt, MRCI+Q, CASPT2, as well as strongly- and partially-contracted NEVPT2. We use MRCI+Q as the benchmark and plot the error in energy ($\Delta E = E - E[\mathrm{MRCI\!+\!Q}]$) along the ground-state potential energy curves (PECs) in \cref{fig:H2O,fig:N2}. We first compare performance of \tdpt with respect to CASPT2. For \ce{H2O}, PECs of both methods exhibit similar mean absolute deviations (\mad) relative to MRCI+Q (\cref{fig:H2O}), while the \tdpt curve is more parallel, as demonstrated by the smaller non-parallelity error (\npe = $\mathrm{max}(E) - \mathrm{min}(E)$ = 4.5 \meh), compared to CASPT2 (\npe = 7.3 \meh). In the case of \ce{N2} (\cref{fig:N2}), this situation is reversed, with CASPT2 giving the smaller \mad errors than \tdpt (\mad = 11.8 and 34.0 \meh, respectively), while the \npe errors for both methods are similar (9.7 and 10.4 \meh).

\begin{table*}[t]
\begin{flushleft}
\captionsetup{justification=raggedright,singlelinecheck=false}
\caption{Total energy (E + 76.0, in \eh) for the symmetric dissociation of water as a function of the O--H bond length (R, \AA). All methods employed the (6e, 9o) CASSCF reference wavefunction and the \mbox{cc-pVQZ} basis set. The H--O--H angle was fixed at 104.5\degree.}
\label{tab:H2O}
{
\setstretch{1.0}
\begin{tabular}{C{1.5cm} C{2.8cm} C{2.8cm} C{2.8cm} C{2.8cm} C{2.8cm}}
\hline
\hline
 \multicolumn{1}{c}{R, \AA} &  \multicolumn{1}{c}{\mrci} &  \multicolumn{1}{c}{\caspt} &  \multicolumn{1}{c}{\sc} &  \multicolumn{1}{c}{\pc} &  \multicolumn{1}{c}{\tdpt} \\
\hline
0.8	& $-$0.32231	& $-$0.30424	& $-$0.30584	& $-$0.30683	& $-$0.30685 \\
0.9	& $-$0.38528	& $-$0.36722	& $-$0.36799	& $-$0.36920	& $-$0.36924 \\
1.0	& $-$0.38902	& $-$0.37094	& $-$0.37158	& $-$0.37305	& $-$0.37309 \\
1.1	& $-$0.36348	& $-$0.34540	& $-$0.34634	& $-$0.34807	& $-$0.34812 \\
1.2	& $-$0.32479	& $-$0.30684	& $-$0.30813	& $-$0.31018	& $-$0.31022 \\
1.3	& $-$0.28175	& $-$0.26418	& $-$0.26536	& $-$0.26774	& $-$0.26780 \\
1.4	& $-$0.23886	& $-$0.22233	& $-$0.21924	& $-$0.22212	& $-$0.22217 \\
1.5	& $-$0.19900	& $-$0.18321	& $-$0.17842	& $-$0.18146	& $-$0.18151 \\
1.6	& $-$0.16342	& $-$0.14836	& $-$0.14311	& $-$0.14610	& $-$0.14615 \\
1.7	& $-$0.13263	& $-$0.11834	& $-$0.11277	& $-$0.11568	& $-$0.11573 \\
1.8	& $-$0.10676	& $-$0.09326	& $-$0.08752	& $-$0.09029	& $-$0.09034 \\
1.9	& $-$0.08570	& $-$0.07292	& $-$0.06715	& $-$0.06976	& $-$0.06981 \\
2.0	& $-$0.06910	& $-$0.05693	& $-$0.05128	& $-$0.05371	& $-$0.05376 \\
2.1	& $-$0.05645	& $-$0.04473	& $-$0.03931	& $-$0.04158	& $-$0.04163 \\
2.2	& $-$0.04709	& $-$0.03570	& $-$0.03056	& $-$0.03271	& $-$0.03275 \\
2.3	& $-$0.04035	& $-$0.02919	& $-$0.02435	& $-$0.02637	& $-$0.02641 \\
2.4	& $-$0.03558	& $-$0.02456	& $-$0.02001	& $-$0.02193	& $-$0.02196 \\
2.5	& $-$0.03224	& $-$0.02131	& $-$0.01701	& $-$0.01883	& $-$0.01887 \\
2.6	& $-$0.02990	& $-$0.01903	& $-$0.01493	& $-$0.01668	& $-$0.01672 \\
2.7	& $-$0.02826	& $-$0.01742	& $-$0.01348	& $-$0.01518	& $-$0.01522 \\
2.8	& $-$0.02710	& $-$0.01628	& $-$0.01247	& $-$0.01413	& $-$0.01416 \\
\hline
\hline
\end{tabular}
}
\end{flushleft}
\end{table*}

\begin{table*}[t]
\begin{flushleft}
\captionsetup{justification=raggedright,singlelinecheck=false}
\caption{Total energy (E + 109.0, in \eh) of \ce{N2} as a function of the N--N bond length (R, \AA). All methods employed the (10e, 10o) CASSCF reference wavefunction and the \mbox{cc-pVQZ} basis set.}
\label{tab:N2}
{
\setstretch{1.0}
\begin{tabular}{C{1.5cm} C{2.8cm} C{2.8cm} C{2.8cm} C{2.8cm} C{2.8cm}}
\hline
\hline
 \multicolumn{1}{c}{R, \AA} &  \multicolumn{1}{c}{\mrci} &  \multicolumn{1}{c}{\caspt} &  \multicolumn{1}{c}{\sc} &  \multicolumn{1}{c}{\pc} &  \multicolumn{1}{c}{\tdpt} \\
\hline
0.9	& 	$-$0.28956 & $-$0.27199	& $-$0.25600 &$-$0.25900	&  $-$0.25936 \\
1.0	& 	$-$0.43066 & $-$0.41272	& $-$0.39751 &$-$0.40049	&  $-$0.40084\\
1.1	& 	$-$0.46373 & $-$0.44560	& $-$0.43088 &$-$0.43386	&  $-$0.43421\\
1.2	& 	$-$0.44276 & $-$0.42467	& $-$0.41012 &$-$0.41314	&  $-$0.41349\\
1.3	& 	$-$0.39769 & $-$0.37989	& $-$0.36513 &$-$0.36824	&  $-$0.36860\\
1.4	& 	$-$0.34495 & $-$0.32775	& $-$0.31235 &$-$0.31559	&  $-$0.31597\\
1.5	& 	$-$0.29323 & $-$0.27699	& $-$0.26046 &$-$0.26388	&  $-$0.26427\\
1.6	& 	$-$0.24689 & $-$0.23201	& $-$0.21382 &$-$0.21743	&  $-$0.21783\\
1.7	& 	$-$0.20792 & $-$0.19480	& $-$0.17441 &$-$0.17817	&  $-$0.17859\\
1.8	& 	$-$0.17703 & $-$0.16575	& $-$0.14268 &$-$0.14649	&  $-$0.14695\\
1.9	& 	$-$0.15385 & $-$0.14369	& $-$0.11270 &$-$0.11691	&  $-$0.11746\\
2.0	& 	$-$0.13759 & $-$0.12873	& $-$0.09558 &$-$0.09968	&  $-$0.10020\\
2.1	& 	$-$0.12673 & $-$0.11832	& $-$0.08447 &$-$0.08825	&  $-$0.08871\\
2.2	& 	$-$0.11970 & $-$0.11121	& $-$0.07734 &$-$0.08080	&  $-$0.08118\\
2.3	& 	$-$0.11521 & $-$0.10637	& $-$0.07285 &$-$0.07602	&  $-$0.07633\\
2.4	& 	$-$0.11234 & $-$0.10307	& $-$0.07004 &$-$0.07296	&  $-$0.07323\\
2.5	& 	$-$0.11048 & $-$0.10078	& $-$0.06827 &$-$0.07100	&  $-$0.07122\\
2.6	& 	$-$0.10924 & $-$0.09918	& $-$0.06715 &$-$0.06972	&  $-$0.06991\\
2.7	& 	$-$0.10839 & $-$0.09804	& $-$0.06643 &$-$0.06887	&  $-$0.06903\\
2.8	& 	$-$0.10779 & $-$0.09721	& $-$0.06596 &$-$0.06829	&  $-$0.06843\\
2.9	& 	$-$0.10736 & $-$0.09661	& $-$0.06565 &$-$0.06789	&  $-$0.06801\\
\hline
\hline
\end{tabular}
}
\end{flushleft}
\end{table*}

We now compare the performance of \tdpt with sc- and pc-NEVPT2. Since the \tdpt results are uncontracted, the difference of the \tdpt and NEVPT2 energies allows us to observe the errors of the internal contraction approximation. These errors are plotted in the inset of \cref{fig:H2O,fig:N2}. The effect of strong contraction in sc-NEVPT2 amounts to errors of 1 to 5 \meh in correlation energy for the systems studied, with significant deviations from parallelity (2 to 3 \meh) relative to \tdpt. The errors of internal contraction in pc-NEVPT2 are much smaller: 0.04 and 0.42 \meh for \ce{H2O} and \ce{N2} respectively, on average. Nevertheless, in the case of \ce{N2}, the pc-NEVPT2 energies exhibit a noticeable non-parallelity error ($\sim$ 0.7 \meh), relative to \tdpt, with the largest deviation of $\sim$ 0.9 \meh at 1.88 \AA, where both methods show steep increase of their errors with respect to MRCI+Q (\cref{fig:N2_zoomed}). 

\subsection{Ground and excited states in \ce{C2}}
\begin{figure*}[t]
   \subfloat[]{\label{fig:C2_dmrg_pec}\includegraphics[width=0.45\textwidth]{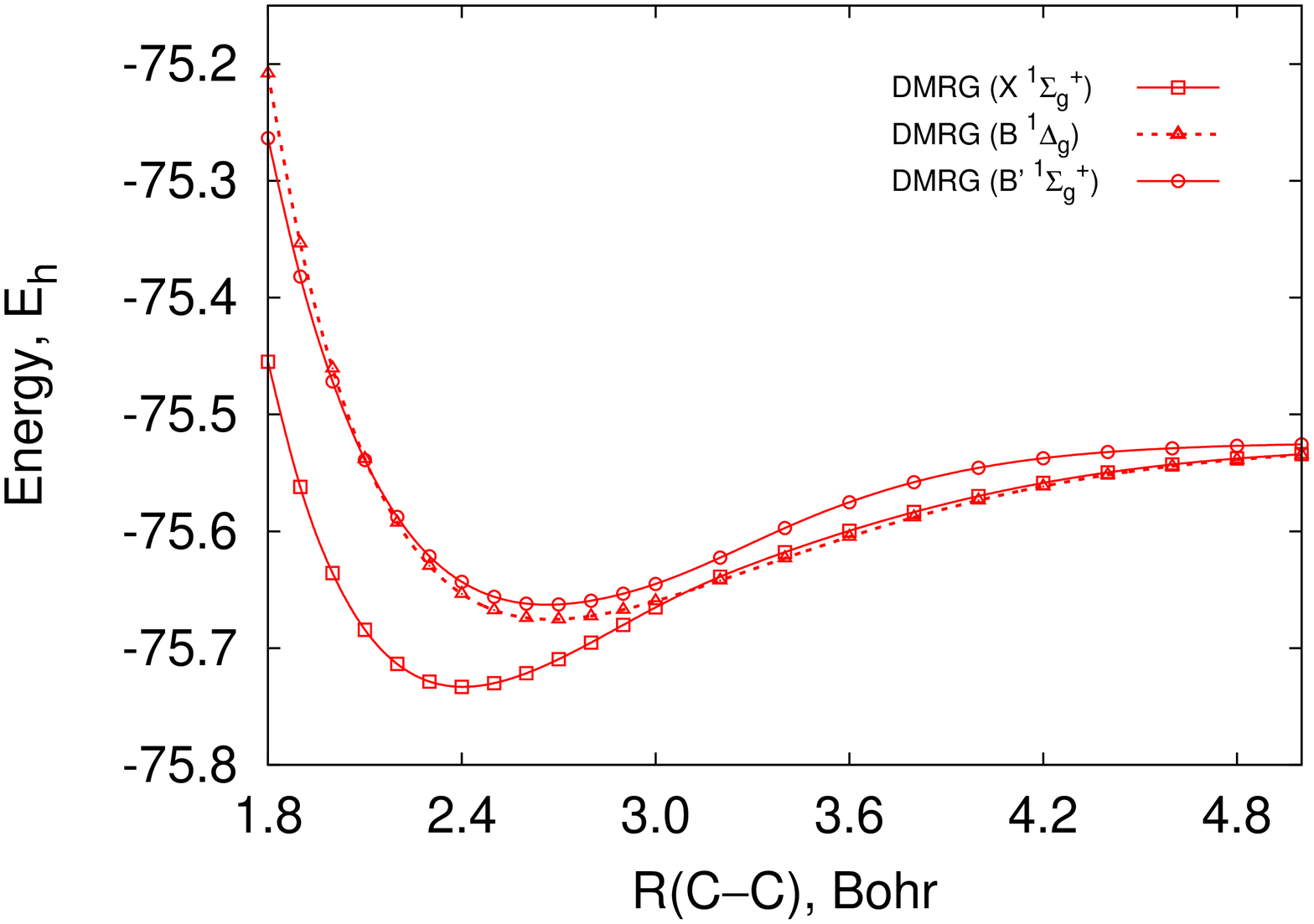}} \qquad
   \subfloat[]{\label{fig:C2_tdpt_pec}\includegraphics[width=0.45\textwidth]{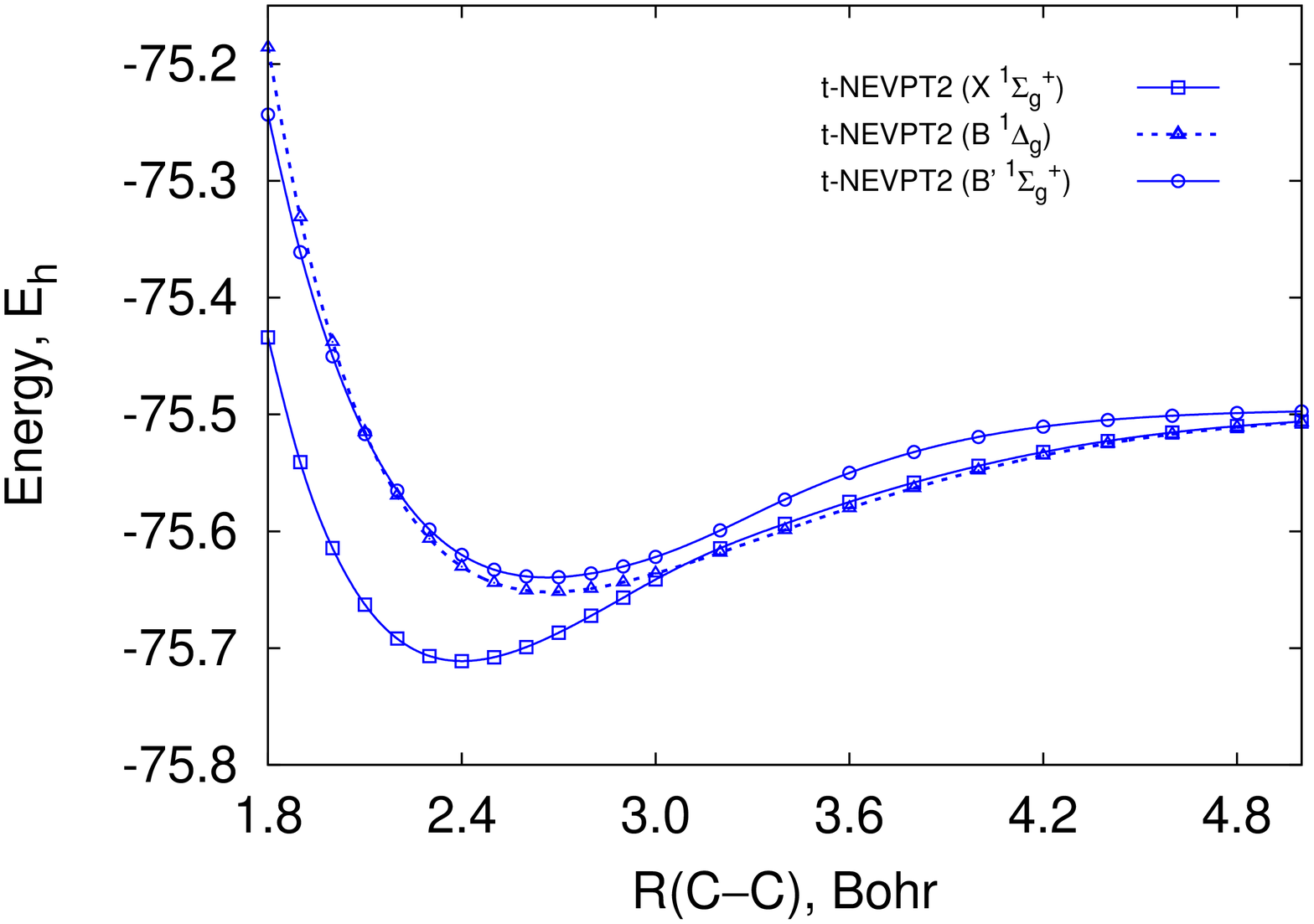}}
   \captionsetup{justification=raggedright,singlelinecheck=false}
   \caption{Potential energy curves for the three singlet states of \ce{C2} computed using (a) DMRG and (b) \tdpt (cc-pVDZ basis set). For \tdpt, the (8e, 8o) CASSCF reference wavefunction was used, with all three states averaged with equal weights. The DMRG results employing the (12e, 28o) active space were obtained from Ref.\@ \citenum{Wouters:2014p1501}.}
   \label{fig:C2_pec}
\end{figure*}

\begin{figure}[t]
   \includegraphics[width=0.45\textwidth]{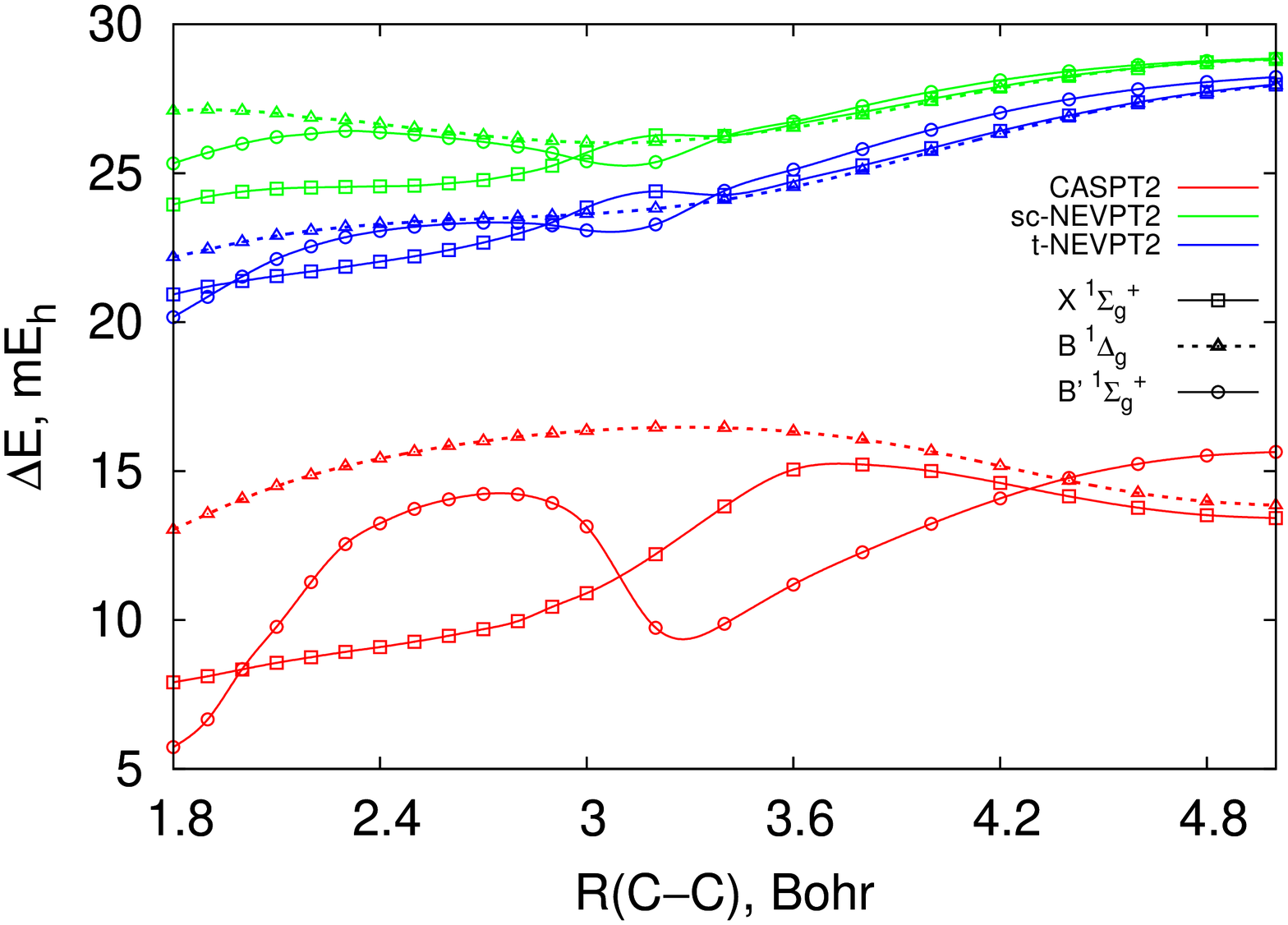}
   \captionsetup{justification=raggedright,singlelinecheck=false}
   \caption{Errors in total energies ($\Delta$E, \meh) of the three singlet states of \ce{C2}, relative to DMRG. For CASPT2, sc-NEVPT2, and \tdpt, the (8e, 8o) CASSCF reference wavefunction was used, with all three states averaged with equal weights. The DMRG results employing the (12e, 28o) active space were taken from Ref.\@ \citenum{Wouters:2014p1501}. All computations employed the cc-pVDZ basis set.}
   \label{fig:C2_diff}
\end{figure}

In this section, we analyze the performance of \tdpt for predicting the properties of excited states in the carbon dimer (\ce{C2}). Specifically, we consider the three singlet electronic states of \ce{C2}: the ground \xsigma state and the two low-lying excited states (\bdelta and \bsigma). For each state, reference wavefunctions were obtained from the state-averaged CASSCF computation using the (8e, 8o) active space, with all three states averaged with equal weights. We employ the recent density matrix renormalization group (DMRG) results by Wouters {\it et al.}\cite{Wouters:2014p1501} as the benchmark. \cref{fig:C2_pec} compares PECs of the three states computed using DMRG and \tdpt, while the errors of CASPT2, sc-NEVPT2, and \tdpt, relative to DMRG, are shown in \cref{fig:C2_diff}. Out of the three methods considered, sc-NEVPT2 exhibits the smallest non-parallelity errors (\npe), while \npe of \tdpt is intermediate between that of sc-NEVPT2 and CASPT2. Both \tdpt and sc-NEVPT2 show a more consistent performance across PECs of different electronic states than CASPT2, which can be observed by comparing the spread of the error curves in \cref{fig:C2_diff} for each method. As a result, the \tdpt and sc-NEVPT2 vertical excitation energies are in significantly better agreement with DMRG, compared to CASPT2. At R(C--C) = 2.4 $a_0$ (near equilibrium), the errors in the ($\bdelta\leftarrow\xsigma$; $\bsigma\leftarrow\xsigma$) vertical excitation energies are (0.03; 0.03), (0.06; 0.05), and (0.17; 0.11) eV for \tdpt, sc-NEVPT2, and CASPT2, respectively. In addition, \tdpt and sc-NEVPT2 show much smaller non-parallelity errors than CASPT2 for the description of avoided crossing of the \xsigma and \bsigma states (2.9 -- 3.4 $a_0$).

\subsection{Chromium dimer}

\begin{figure}[t]
   \subfloat[]{\label{fig:Cr2_nevpt_pec}\includegraphics[width=0.45\textwidth]{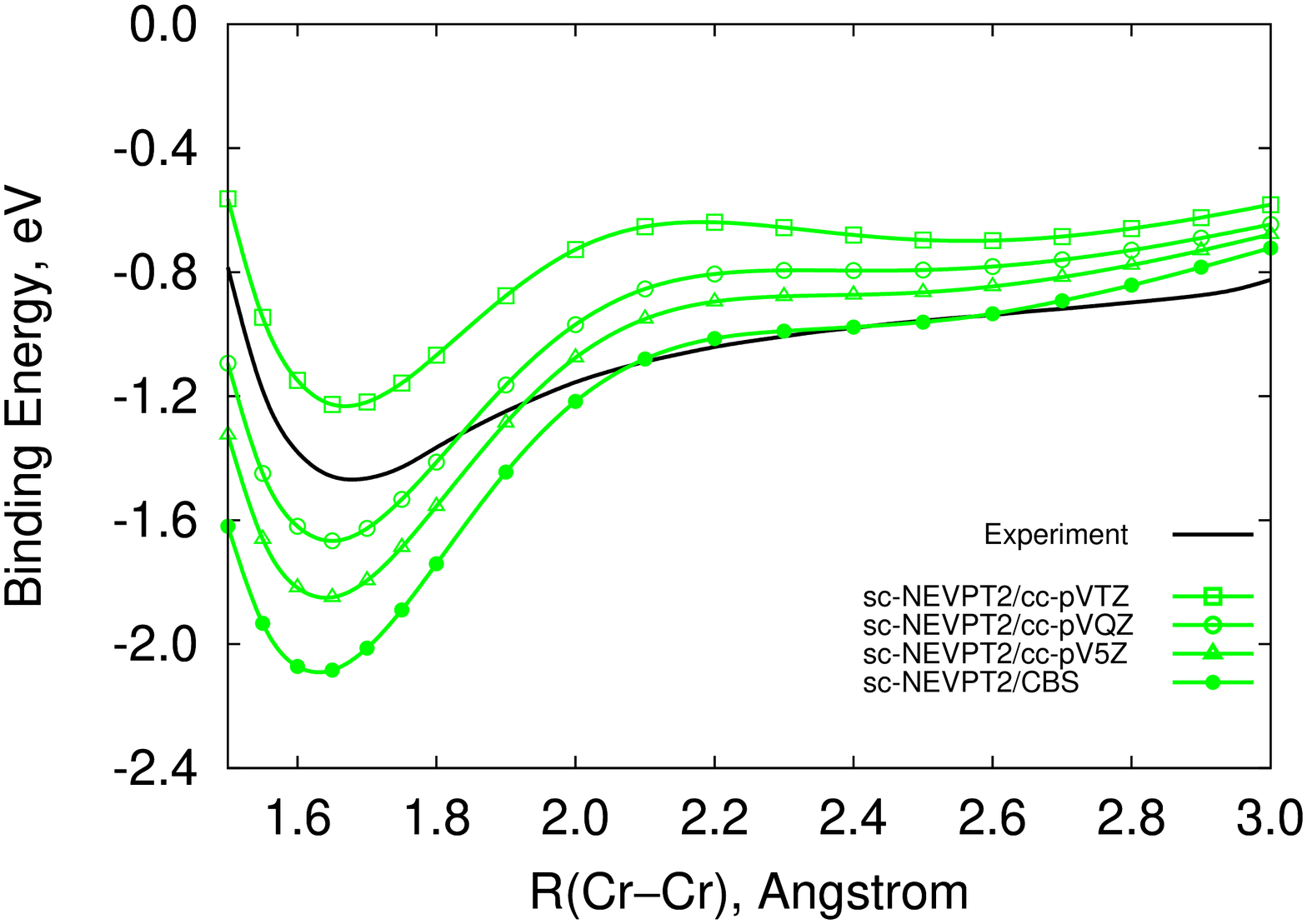}} \quad
   \subfloat[]{\label{fig:Cr2_tdpt_pec}\includegraphics[width=0.45\textwidth]{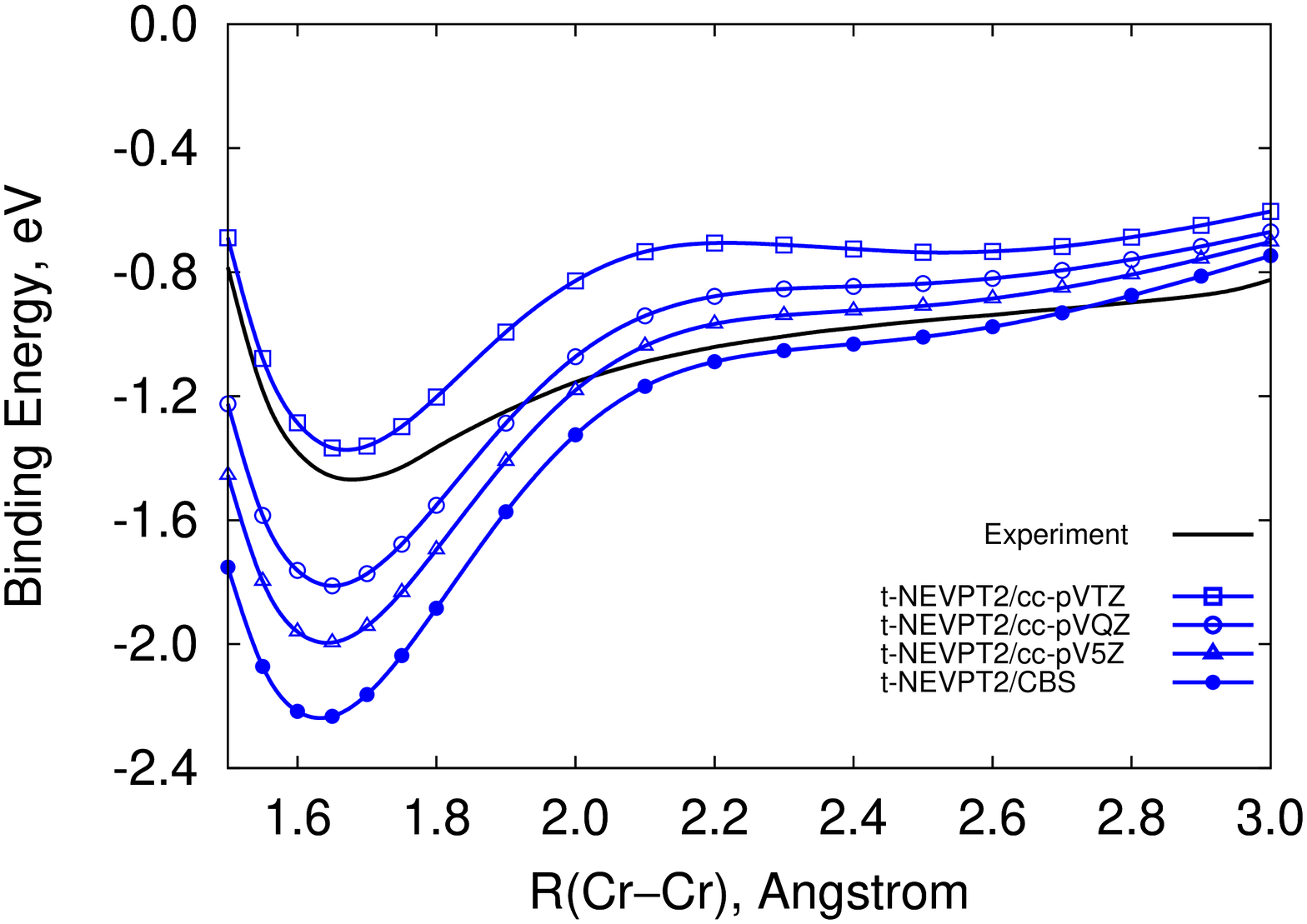}}
   \captionsetup{justification=raggedright,singlelinecheck=false}
   \caption{Ground-state potential energy curves (in eV) for the dissociation of \ce{Cr2} computed using (a) sc-NEVPT2 and (b) \tdpt. Both methods employed the (12e, 12o) CASSCF reference wavefunction and the cc-pVXZ basis sets (X = T, Q, 5). Energies are relative to the isolated atoms. The complete basis set limit (CBS) results were obtained as discussed in Sec.\@ \ref{sec:compdetails}. Experimental curve is taken from Ref.\@ \citenum{Casey:1993p816}. }
   \label{fig:Cr2_pec}
\end{figure}

Finally, we turn our attention to the chromium dimer, whose
ground-state PEC is notoriously difficult
to describe well theoretically. The \ce{Cr2} molecule has been the subject of many computational studies.\cite{Andersson:1994p391,Roos:1995p215,Andersson:1995p212,Stoll:1996p793,Dachsel:1999p152,Roos:2003p265,Celani:2004p2369,Angeli:2006p054108,Ruiperez:2011p1640,Muller:2009p12729,Kurashige:2011p094104,Purwanto:2015p064302,Guo:2015arXiv} It has been shown that, for a proper description of the \ce{Cr2} dissociation curve, a combination of high levels of theory with very large basis sets is required. With respect to the choice of multi-reference methodology, accurate results have recently been obtained using the
multi-reference-averaged quadratic coupled cluster method,\cite{Muller:2009p12729} auxiliary-field quantum Monte Carlo,\cite{Purwanto:2015p064302} and the combination of the internally-contracted perturbation theories (CASPT2 and NEVPT2) with DMRG.\cite{Kurashige:2011p094104,Guo:2015arXiv} The latter perturbation
theory studies have demonstrated that multi-reference wavefunctions based on the valence (12e, 12o) active space, which contains the $3d$ and $4s$ orbitals of chromium atoms, generally do not provide a sufficiently good zeroth-order approximation for perturbative treatments,\cite{Celani:2004p2369,Angeli:2006p054108,Ruiperez:2011p1640} and including  extra $4d$ shells is necessary to achieve quantitative agreement with the experiment.\cite{Kurashige:2011p094104,Guo:2015arXiv}
However, an open question is the effect of contraction in the perturbation theory treatment of the \ce{Cr2} curve.
Here, we  perform the first uncontracted perturbation theory study of the \ce{Cr2} PEC by employing our \tdpt algorithm. 
As we currently can only use CASSCF reference wavefunctions, 
we will limit our analysis to the (12e, 12o) active space, and investigate the effect of the strong contraction approximation by comparing the
results of \tdpt with sc-NEVPT2 in the same active space.

\begin{figure}[t]
   \includegraphics[width=0.45\textwidth]{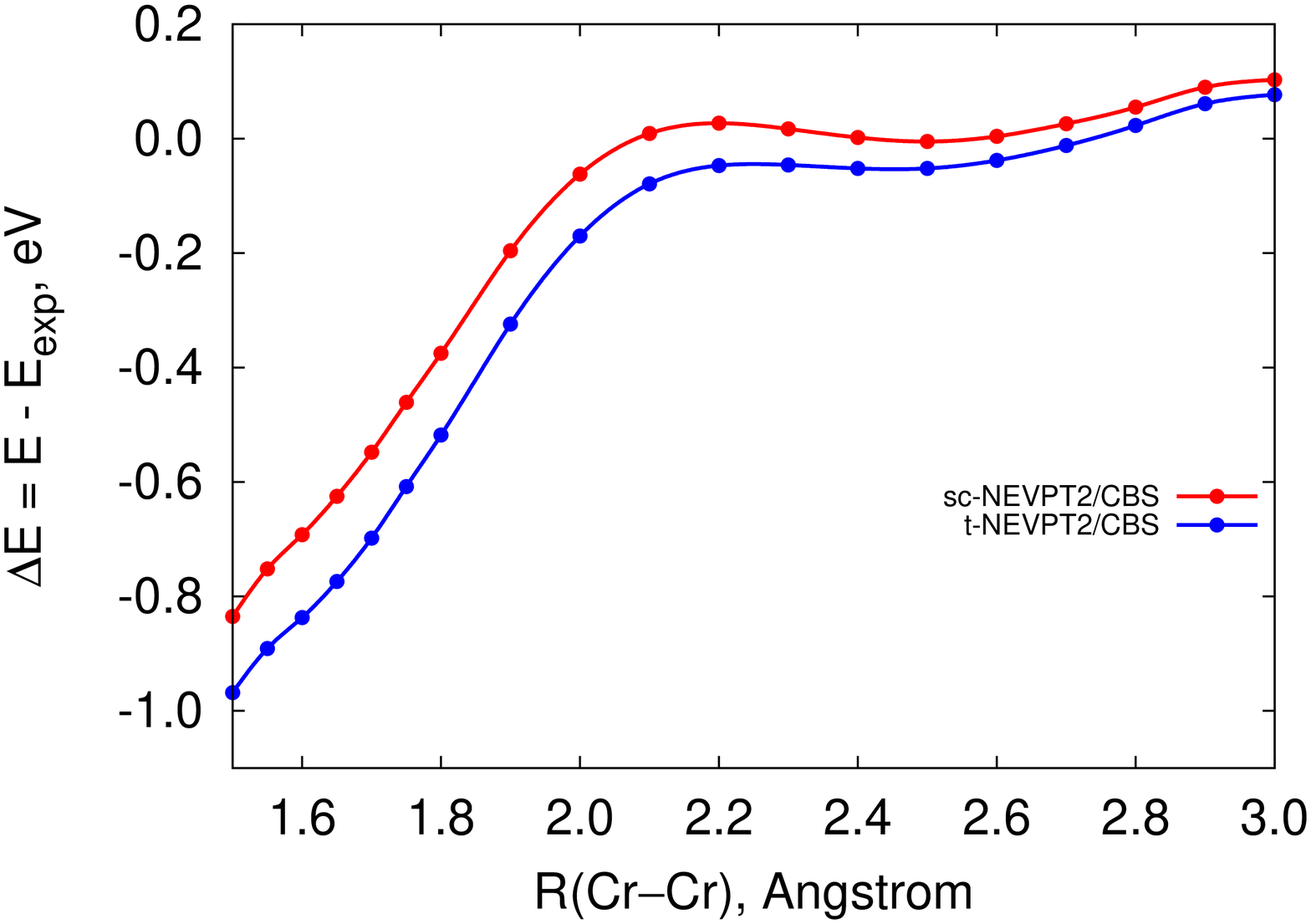}
   \captionsetup{justification=raggedright,singlelinecheck=false}
   \caption{Ground-state binding energies (in eV) for the dissociation of \ce{Cr2} computed using sc-NEVPT2 and \tdpt, relative to experiment.\cite{Casey:1993p816} Energies were extrapolated to the complete basis set limit. Both methods employed the (12e, 12o) CASSCF reference wavefunction.}
   \label{fig:Cr2_diff}
\end{figure}

\cref{fig:Cr2_pec} shows the ground-state PECs for the \ce{Cr2} dissociation curve computed using the cc-pV$X$Z basis sets ($X$ = T, Q, 5) and extrapolated to the complete basis set limit (CBS). Using the small (12e, 12o) active space, both sc-NEVPT2 and \tdpt significantly overbind (by $\sim$ 0.9 eV) around the equilibrium region (1.5 -- 2.1 \AA), relative to experiment.\cite{Casey:1993p816} Also, as previously seen, the computed PECs strongly depend on the basis set,  giving rise to the large correction of the CBS extrapolation. When using the cc-pVTZ basis set, both perturbative methods produce an incorrect shape of
the PECs, with a barrier in the range of 2.0 -- 2.3 \AA. However, in the case of \tdpt, the magnitude of the barrier is significantly lowered compared to sc-NEVPT2. Increasing the basis set results in the disappearance of the barrier and the appearance of the characteristic ``shoulder'' on the PECs (2.1 -- 2.8 \AA). In this region, the CBS-extrapolated PECs for both methods exhibit a good agreement with experiment, with errors of less than 0.1 eV. \cref{fig:Cr2_diff} compares the sc-NEVPT2 and \tdpt binding energies relative to experiment. The effect of the strong contraction approximation
amounts to $\sim$ 0.15 eV in \ce{Cr2} dissociation energy in the equilibrium region and decreases monotonically towards the dissociation limit.
Interestingly, the recent results of Guo {\it et al.}\cite{Guo:2015arXiv} using sc-NEVPT2 on top of a DMRG reference wavefunction in a (12e, 22o) active space (including the $4d$ shells and relativistic effects) yields a slight underbinding relative to experiment, by about 0.05 eV. Assuming
a smaller effect of contraction in the larger active space employed by Guo {\it et al.}, the underbinding may be corrected by using the
uncontracted formulation.
Finally,  as  seen from \cref{fig:Cr2_diff}, in the shoulder region the uncontracted \tdpt PEC is slightly more parallel to experiment
than that of sc-NEVPT2.

\section{Conclusions}

\label{sec:conclusions}

In this work we argued for considering a time-dependent implementations of multi-reference perturbation theory. The essential reason is
that it is complicated to represent the resolvent operator for a complicated zeroth-order Hamiltonian, while
it is simple to represent the corresponding time-evolution of the same operator. 
As an example, we provided the formulation and implementation of
second-order perturbation theory using the Dyall Hamiltonian. The corresponding theory is equivalent to fully uncontracted
$n$-electron valence perturbation theory, but reduces the scaling relative to contracted variants of the theory, particularly
with respect to the number of active orbitals ($\mathcal{O}(N_{det} \times N_{act}^6)$ versus $\mathcal{O}(N_{det} \times N_{act}^8)$ scaling) and further avoids the
need to diagonalize large metric matrices. Using this formulation we examined the effect of contraction in multi-reference perturbation theory
in a variety of model problems, including water, nitrogen dimer, carbon dimer, and the chromium dimer, in large and realistic basis sets.

Several extensions of the current work can be imagined. An immediate extension is to combine the time-dependent formulation
of the $n$-electron valence perturbation theory with DMRG reference wavefunctions, thus providing an uncontracted analogue
of the recently reported combination of strongly-contracted $n$-electron valence perturbation theory with the DMRG reference.\cite{Guo:2015arXiv}
Such an uncontracted theory will also be closely related to the recently reported MPS-PT2 theory, although it would use
the time-dependent DMRG algorithm, rather than relying on MPS compression of the Hylleraas functional as in MPS-PT2.\cite{Sharma:2014p111101} There, the efficiency and accuracy of the DMRG compression during imaginary-time evolution remains to be studied. Alternatively,
the generalization of higher-order diagrammatic perturbation theories and resummations using interacting zeroth-order Hamiltonians is
a natural extension of this work. In this context, comparison with the recent renormalized Green's function techniques based on impurity
formulations, which incorporate different kinds of contributions into the diagrammatic expansion, would appear interesting and fruitful.\cite{Phillips:2014p241101,Lan:2015p241102}

\section{Acknowledgements}
This work was supported by the US Department of Energy, Office of Science through
Award DE-SC0008624 (primary support for A.Y.S.), and Award DE-SC0010530 (additional support for G.K.-L.C.). A.Y.S. would like to thank Dr.\@ Qiming Sun, Dr.\@ Enrico Ronca and Sheng Guo for insightful discussions.


\section{Appendix: $\mathbf{t}$-NEVPT2 Energy Contributions}
Components of the perturbation operator $\hat{V}^\prime$ (\cref{eq:v_prime}) are defined as:
\begin{align}
	\label{eq:v_prime_explicit}
	\hat{V}^{[0]} &= \frac{1}{4} \sum_{ijab} \v{ab}{ij} \c{a} \c{b} \a{j} \a{i} \ , \\
	\hat{V}^{[+1]} &= \frac{1}{2} \sum_{ijxa} \v{ax}{ij} \c{a} \c{x} \a{j} \a{i} \ , \\
	\hat{V}^{[-1]} &= \frac{1}{2} \sum_{ixab} \v{ab}{ix} \c{a} \c{b} \a{x} \a{i} \ , \\
	\hat{V}^{[+2]} &= \frac{1}{4} \sum_{ijxy} \v{xy}{ij} \c{x} \c{y} \a{j} \a{i} \ , \\
	\hat{V}^{[-2]} &= \frac{1}{4} \sum_{xyab} \v{ab}{xy} \c{a} \c{b} \a{y} \a{x} \ , \\
	\hat{V}^{[+1']} &= \sum_{ix} \h{x}{i} \c{x} \a{i} + \frac{1}{2} \sum_{ijx} \v{xj}{ij} \c{x} \c{j} \a{j} \a{i} \notag \\ 
	&+ \frac{1}{2} \sum_{ixyz} \v{yz}{ix} \c{y} \c{z} \a{x} \a{i} \ , \\
	\hat{V}^{[-1']} &= \sum_{xa} \h{a}{x} \c{a} \a{x} + \sum_{ixa} \v{ai}{xi} \c{a} \c{i} \a{i} \a{x} \notag \\ 
	&+ \frac{1}{2} \sum_{xyaz} \v{az}{xy} \c{a} \c{z} \a{y} \a{x} \ , \\
	\hat{V}^{[0']}  &= \sum_{ia} \h{a}{i} \c{a} \a{i} + \frac{1}{2} \sum_{ija} \v{aj}{ij} \c{a} \c{j} \a{j} \a{i} \notag \\ 
	&+ \sum_{ixya} \v{ay}{ix} \c{a} \c{y} \a{x} \a{i} \ .
\end{align}
\begin{widetext}
The \tdpt correlation energy contributions (\cref{eq:e_contributions}) can be written as:
\begin{align}
	\label{eq:e_0}
	E^{[0]} &= \frac{1}{4} \sum_{ijab} \frac{\v{ij}{ab} \v{ab}{ij}}{\e{i} + \e{j} - \e{a} - \e{b}} \ , \\
	\label{eq:e_+1}
	E^{[+1]} &= -\frac{1}{2} \sum_{xy} \int_{0}^{\infty} \sum_{ija} \v{ij}{ay} \v{ax}{ij} e^{(\e{i} + \e{j} - \e{a}) \tau} \braket{\Psi_0| \a{y} (\tau) \c{x} |\Psi_0} \mathrm{d}\tau \ , \\
	\label{eq:e_-1}
	E^{[-1]} &= -\frac{1}{2} \sum_{xy} \int_{0}^{\infty}  \sum_{iab} \v{iy}{ab} \v{ab}{ix} e^{(\e{i} - \e{a} - \e{b}) \tau} \braket{\Psi_0| \c{y} (\tau) \a{x} |\Psi_0} \mathrm{d}\tau \ , \\
	\label{eq:e_+2}
	E^{[+2]} &= -\frac{1}{8} \sum_{wxyz} \int_{0}^{\infty}  \sum_{ij} \v{ij}{zw} \v{xy}{ij} e^{(\e{i} + \e{j}) \tau} \braket{\Psi_0| \a{z} (\tau) \a{w} (\tau) \c{y} \c{x} |\Psi_0} \mathrm{d}\tau \ , \\
	\label{eq:e_-2}
	E^{[-2]} &= -\frac{1}{8} \sum_{wxyz} \int_{0}^{\infty}  \sum_{ab} \v{zw}{ab} \v{ab}{xy} e^{-(\e{a} + \e{b}) \tau} \braket{\Psi_0| \c{z} (\tau) \c{w} (\tau) \a{y} \a{x} |\Psi_0} \mathrm{d}\tau \ , \\
	\label{eq:e_+1p}
	E^{[+1']} 
	&= -\sum_{xy} \int_{0}^{\infty}  \sum_{i} \th{i}{y} \th{x}{i} e^{\e{i} \tau} \braket{\Psi_0| \a{y} (\tau) \c{x} |\Psi_0} \mathrm{d}\tau \notag \\
	&-  \sum_{wxyz} \int_{0}^{\infty}  \sum_{i} \th{i}{w} \v{zy}{ix} e^{\e{i} \tau} \braket{\Psi_0| \a{w} (\tau) \c{z} \c{y} \a{x}  |\Psi_0} \mathrm{d}\tau \notag  \\
	&-   \frac{1}{4} \sum_{\substack{uvw\\xyz}} \int_{0}^{\infty} \sum_{i} \v{iu}{wv} \v{zy}{ix} e^{\e{i} \tau} \braket{\Psi_0| \c{u} (\tau) \a{v} (\tau) \a{w} (\tau) \c{z} \c{y} \a{x} |\Psi_0} \mathrm{d}\tau \ , \\
	\label{eq:e_-1p}
	E^{[-1']}
	&= -\sum_{xy} \int_{0}^{\infty}  \sum_{a} \th{y}{a} \th{a}{x} e^{-\e{a} \tau} \braket{\Psi_0| \c{y} (\tau) \a{x} |\Psi_0} \mathrm{d}\tau \notag \\
	&- \sum_{wxyz} \int_{0}^{\infty}  \sum_{a} \th{w}{a} \v{az}{xy} e^{-\e{a} \tau} \braket{\Psi_0| \c{w} (\tau) \c{z} \a{y} \a{x} |\Psi_0} \mathrm{d}\tau \notag  \\
	&- \frac{1}{4} \sum_{\substack{uvw\\xyz}} \int_{0}^{\infty}  \sum_{a} \v{uv}{aw} \v{az}{xy} e^{-\e{a} \tau} \braket{\Psi_0| \c{u} (\tau) \c{v} (\tau) \a{w} (\tau) \c{z} \a{y} \a{x} |\Psi_0} \mathrm{d}\tau \ , \\
	\label{eq:e_0p}
	E^{[0']} &= \sum_{ia} \frac{\th{i}{a} \th{a}{i}}{\e{i} - \e{a}}
	+ 2 \sum_{ixya} \frac{\th{a}{i} \v{ix}{ay} \pdm{y}{x} }{\e{i} - \e{a}} \notag \\
	&- \sum_{wxyz} \int_{0}^{\infty} \sum_{ia} \v{iz}{aw} \v{ay}{ix} e^{(\e{i} - \e{a})\tau} \braket{\Psi_0| \c{z} (\tau) \a{w} (\tau) \c{y} \a{x} |\Psi_0} \mathrm{d}\tau \ .
\end{align}
In \cref{eq:e_0,eq:e_+1,eq:e_-1,eq:e_+2,eq:e_-2,eq:e_+1p,eq:e_-1p,eq:e_0p}, the $\th{p}{q}$ matrix elements are defined as:
\begin{align}
	\label{eq:h_core}
	\th{p}{q} = \h{p}{q} + \sum_{i} \v{pi}{qi} \ .
\end{align}
\end{widetext}

\bibliographystyle{jcp}

\end{document}